\documentclass[10pt,aps,prd,twocolumn,showpacs,amsmath,amssymb,nofootinbib,eqsecnum,preprintnumbers,longbibliography]{revtex4-1}

\usepackage[usenames,dvipsnames]{xcolor}	%% colored notes
\usepackage{mathtools}						%% mathclap, etc.
\usepackage{tensor}							%% tensor indices
\usepackage{suffix} 						%% newcommand with *
\usepackage{xargs} 							%% newcommandx
\usepackage{relsize} 						%% mathsmaller, etc.
\usepackage[mathscr]{eucal} 				%% math fonts

\usepackage{stmaryrd}						%% \varovee symbol
\SetSymbolFont{stmry}{bold}{U}{stmry}{m}{n}
	
\usepackage[english]{babel}					%% longbibliography

\newcommand\bs[1]{\boldsymbol{#1}}

\newcommand\dd{\mathrm{d}}
\newcommand\pp{\partial}

\newcommand\const{\mathrm{const.}}

\renewcommandx{\t}[3][1={}, 3={}]{\tensor[#1]{#2}{^{\mathstrut}_{\mathstrut}#3}}
\newcommandx{\tb}[3][1={}, 3={}]{\t[#1]{}\mkern-1mu\bs{#2}\mkern-1mu\t{}[#3]}

\WithSuffix\newcommandx\t*[3][1={}, 3={}]{\tensor*[#1]{#2}{^{\mathstrut}_{\mathstrut}#3}}
\WithSuffix\newcommandx\tb*[3][1={}, 3={}]{\t*[#1]{}\mkern-1mu\bs{#2}\mkern-1mu\t*{}[#3]}

\newcommand{\nablai}{\nabla\!\mkern-1mu}

\newcommand{\ind}[1]{\mathsmaller{#1}\mkern-1mu}

\newcommandx{\tvec}[2][2={}]{\frac{\smash{\tb{\pp}[#2]}}{\bs{\pp}#1}}

\newcommandx{\ev}[2][2={}]{\t*{\bs{e}}[^{#2}_{#1}]}
\newcommandx{\ef}[2][2={}]{\t*{\bs{e}}[^{#1}_{#2}]}
\newcommandx{\epsv}[2][2={}]{\t*{\bs{\epsilon}}[^{#2}_{#1}]}
\newcommandx{\epsf}[2][2={}]{\t*{\bs{\epsilon}}[^{#1}_{#2}]}

\newcommandx{\evh}[2][2={}]{\t*{\bs{\hat{e}}}[^{#2}_{#1}]}
\newcommandx{\efh}[2][2={}]{\t*{\bs{\hat{e}}}[^{#1}_{#2}]}
\newcommandx{\epsvh}[2][2={}]{\t*{\bs{\hat\epsilon}}[^{#2}_{#1}]}
\newcommandx{\epsfh}[2][2={}]{\t*{\bs{\hat\epsilon}}[^{#1}_{#2}]}

\newcommandx{\evt}[2][2={}]{\t*{\bs{\tilde{e}}}[^{#2}_{#1}]}
\newcommandx{\eft}[2][2={}]{\t*{\bs{\tilde{e}}}[^{#1}_{#2}]}
\newcommandx{\epsvt}[2][2={}]{\t*{\bs{\tilde{\epsilon}}}[^{#2}_{#1}]}
\newcommandx{\epsft}[2][2={}]{\t*{\bs{\tilde{\epsilon}}}[^{#1}_{#2}]}

\newcommandx{\evht}[2][2={}]{\t*{\bs{\hat{\tilde{e}}}}[^{#2}_{#1}]}
\newcommandx{\efht}[2][2={}]{\t*{\bs{\hat{\tilde{e}}}}[^{#1}_{#2}]}
\newcommandx{\epsvht}[2][2={}]{\t*{\bs{\hat\tilde{\epsilon}}}[^{#2}_{#1}]}
\newcommandx{\epsfht}[2][2={}]{\t*{\bs{\hat\tilde{\epsilon}}}[^{#1}_{#2}]}

\newcommandx{\evb}[2][2={}]{\t*{\bs{\bar{e}}}[^{#2}_{#1}]}
\newcommandx{\efb}[2][2={}]{\t*{\bs{\bar{e}}}[^{#1}_{#2}]}
\newcommandx{\epsvb}[2][2={}]{\t*{\bs{\bar{\epsilon}}}[^{#2}_{#1}]}
\newcommandx{\epsfb}[2][2={}]{\t*{\bs{\bar{\epsilon}}}[^{#1}_{#2}]}

\newcommandx{\evhb}[2][2={}]{\t*{\bs{\hat{\bar{e}}}}[^{#2}_{#1}]}
\newcommandx{\efhb}[2][2={}]{\t*{\bs{\hat{\bar{e}}}}[^{#1}_{#2}]}
\newcommandx{\epsvhb}[2][2={}]{\t*{\bs{\hat\bar{\epsilon}}}[^{#2}_{#1}]}
\newcommandx{\epsfhb}[2][2={}]{\t*{\bs{\hat\bar{\epsilon}}}[^{#1}_{#2}]}

\newcommandx{\eR}[8][2={},4={},6={},8={}]{\big(\ef{#1}[#2]\wedge\ef{#3}[#4]\big)\varovee\big(\ef{#5}[#6]\wedge\ef{#7}[#8]\big)}
\newcommandx{\eRh}[8][2={},4={},6={},8={}]{\big(\efh{#1}[#2]\wedge\efh{#3}[#4]\big)\varovee\big(\efh{#5}[#6]\wedge\efh{#7}[#8]\big)}
\newcommandx{\eRr}[8][2={},4={},6={},8={}]{\big(\ef{#1}[#2]\wedge\ef{#3}[#4]\big)\varovee\big(\efh{#5}[#6]\wedge\efh{#7}[#8]\big)}
\newcommandx{\eRm}[8][2={},4={},6={},8={}]{\big(\ef{#1}[#2]\wedge\efh{#3}[#4]\big)\varovee\big(\ef{#5}[#6]\wedge\efh{#7}[#8]\big)}

\newcommand{\Ric}{\mathrm{Ric}}

\newcommand\imag{i}

\newcommand\op[1]{\mathsf{#1}}

\newcommand\gden{\mathfrak{g}}

\newcommandx\sn[3][3={}]{[#1\,{,}\,#2] \t*{}[_{\mathsmaller{\mathrm{SN}}}^{#3}]}
\newcommand\com[2]{[ #1\,{,}\,#2 ]}
\newcommand\comLR[2]{\left[ #1\,{,}\,#2 \right]}

\newcommand\feq{\mathrel{\phantom{=}}}

\newtheorem{theorem}{Theorem}[section]

\newtheorem{lemma}[theorem]{Lemma}

\begin{document}

\title{Spacetimes with a separable Klein--Gordon equation in higher dimensions}

\author{Ivan Kol\'a\v{r}}
\email{Ivan.Kolar@utf.mff.cuni.cz}
\affiliation{Institute of Theoretical Physics,
Faculty of Mathematics and Physics, Charles University in Prague,
V~Hole\v{s}ovi\v{c}k\'ach 2, 180 00 Prague, Czech Republic}

\author{Pavel Krtou\v{s}}
\email{Pavel.Krtous@utf.mff.cuni.cz}
\affiliation{Institute of Theoretical Physics,
Faculty of Mathematics and Physics, Charles University in Prague,
V~Hole\v{s}ovi\v{c}k\'ach 2, 180 00 Prague, Czech Republic}

\date{September 29, 2015}  % version 2.04 - PRD v2

\begin{abstract}
We study spacetimes that lead to a separable Klein--Gordon equation in a general number of dimensions. We introduce an ansatz for the metric in higher dimensions motivated by analogical work by Carter in four dimensions and find solutions of the Klein--Gordon equation. For such a metric we solve the Einstein equations and regain the Kerr--NUT--(A)dS spacetime as one of our results. Other solutions lead to the Einstein--K\"ahler metric of a Euclidean signature. Next we investigate a warped geometry of two Klein--Gordon separable spaces with a properly chosen warped factor. We show that the resulting metric leads also to a separable Klein--Gordon equation and we find the corresponding solutions. Finally, we solve the Einstein equations for the warped geometry and obtain new solutions.

\end{abstract}

%% 02.30.Ik  Integrable systems
%% 02.40.Yy  Geometric mechanics
%% 45.10.Na  Geometrical and tensorial methods
%% 45.20.Jj  Lagrangian and Hamiltonian mechanics
%% 04.50.-h  Higher-dimensional gravity and other theories of gravity
%% 04.50.Gh  Higher-dimensional black holes, black strings, and related objects
%% 04.40.Nr  Einstein-Maxwell spacetimes, spacetimes with fluids, radiation or classical fields
%% 04.70.Bw  Classical black holes

\maketitle

\section{Introduction}\label{sc:intro}

The Kerr solution in four dimensions is one of the most important solutions of the Einstein equations. It allows us to describe the exterior of a stationary rotating central source. Although it deals with a rather elementary physical configuration, it took a significant effort and rather long time to find it---until 1963, 47 years after the discovery of the spherical Schwarzschild solution \cite{Kerr:1963}.

The Kerr solution is not important only for its astrophysical applications but it is also exceptional for its mathematical structure and high symmetry. Geodesic motion in this spacetime is completely integrable and the main physical field equations can be solved by separation of constants. Exactly these properties are the cornerstones of Carter's rederivation and generalization of the Kerr solution in 1968. In his  seminal paper \cite{Carter:1968b}, cf.\ also \cite{Carter:2009}, he studied spaces allowing a separable Hamilton--Jacobi equation and Klein--Gordon equation (referred to as the Schr\"odinger equation by Carter), which satisfy the Einstein equations. He obtained the Kerr solutions with a cosmological constant and NUT parameter, and some other related spaces. The aim of this paper is to follow this route in a higher-dimensional setting.

Higher dimensional solutions of the Einstein equations have become popular in recent years due to their significance in string theory or in the AdS/CFT context. The study of general relativity in a general number of dimensions is interesting also just from the mathematical point of view, and it can reveal some important properties of various structures and equations. Similarly to the four dimensions, finding a solution describing a generally rotating black hole in higher dimensions was a challenging task. The spherically symmetric Tangherlini--Schwarzschild solution \cite{Tangherlini:1963} was found in the same year as the Kerr solution, in prehistory of the higher-dimensional studies. The solution describing a rotating asymptotically flat black hole was discovered by Myers and Perry \cite{MyersPerry:1986} in 1986. Generalization for a nontrivial cosmological constant has been studied on various levels of generality \cite{HawkingEtal:1999,GibbonsEtal:2004,GibbonsEtal:2005} until 2006 when Chen, L\"u, and Pope were able to add also the so-called NUT parameters and to write down the most general metric, desribing the geometry known today as the \emph{Kerr--NUT--(A)dS spacetime} \cite{ChenLuPope:2006,ChenLuPope:2007}.

It is remarkable that although the derivation of this geometry occured through an unrelated route (the Kerr-Schild form of the metric) the final form of the metric is a straightforward generalization of Carter's four-dimensional Kerr--NUT--(A)dS metric \cite{Carter:1968b}. It turns out also that most of the miraculous properties of the Kerr solution survive in higher dimensions: geodesic motion is completely integrable \cite{PageEtal:2007}, Hamilton--Jacobi, Dirac, and Klein--Gordon equations are separable \cite{FrolovEtal:2007,SergyeyevKrtous:2008,CarigliaEtal:2011a,CarigliaEtal:2011b}, and the geometry possesses a high degree of symmetry encoded in the existence of a tower of Killing vectors and tensors \cite{KrtousEtal:2007a,KrtousEtal:2007b}. These results have completed a series of previous works on integrability and separability of the basic physical equations in black hole spacetimes in particular dimensions and for particular values of spacetime parameters \cite{WalkerPenrose:1970,Teukolsky:1972,Teukolsky:1973,Carter:1977,DemianskiFrancaviglia:1981,FrolovStojkovic:2003a,FrolovStojkovic:2003b,%
ChongEtal:2004,VasudevanEtal:2004,KunduriLucietti:2005,VasudevanEtal:2005,VasudevanStevens:2005,ChenLuPope:2006b,Davis:2006,FrolovKubiznak:2007,KubiznakFrolov:2007}. It turns out that the Kerr--NUT--(A)dS spacetime is a very prominent example of a fully integrable and separable system  \cite{BenentiFrancaviglia:1979,BenentiFrancaviglia:1980}.

It is interesting that all these properties hold for a wider class of metrics---the so-called off-shell Kerr--NUT--(A)dS spaces---obtained by relaxing conditions on some of the metric functions in the Kerr--NUT--(A)dS solution. It was shown that this class is uniquely determined by requiring the existence of a principal closed conformal Killing--Yano form \cite{HouriEtal:2007,KrtousFrolovKubiznak:2008,HouriEtal:2008b,HouriEtal:2009}.

It is natural to ask if the elegant Carter's method could be employed also in higher dimensions and whether it could not lead to a more general class of solutions. In this paper we investigate this possibility. We assume a metric form motivated by a combination of Carter's metric in four dimensions and the Kerr--NUT--(A)dS metric in a general number of dimensions. We show that such a metric leads to a separable Klein--Gordon equation and we find the corresponding solutions. We call this class the Klein--Gordon simple separable metrics.\footnote{%
The class is specified by the ansatz \eqref{eq:zmetric} below. Although all such metrics lead to a separable Klein--Gordon equation, not all metrics leading to a separable Klein--Gordon equation belong to this class, as we will observe below.} Moreover, according to \cite{ChervonyiLunin:2015}, it seems that metrics which yield a separable Hamilton--Jacobi equation should also have this form.

The introduced class generalizes the off-shell Kerr--NUT-(A)dS metrics by introducing one functional reparametrization freedom for each latitude and radial coordinate. It is indeed a wider class, and thus not all Klein--Gordon simple separable metrics possess the (standard) principal closed conformal Killing--Yano form, however, they still allow the tower of Killing tensors and vectors. In fact, the class of simple separable metrics belongs to a wider class of metrics admitting a generalized closed conformal Killing-Yano form with torsion \cite{HouriEtal:2012}, which justifies the exceptional symmetry.

Next, we solve the Einstein equations in the class of Klein--Gordon simple separable metrics. One family of solutions contains exactly the on-shell Kerr--NUT--(A)dS spacetimes. However, we also find another family of solutions, which are special Einstein--K\"ahler metrics. Unfortunately, these metrics necessarily have a Euclidean signature, which cannot be Wick-rotated to the physical sector. Probably, for this reason, a similar solution was not further considered by Carter in four dimensions.

However, in higher dimensions, we have several ways of building a geometry from smaller blocks. One of such methods is a warped product of two metrics. We employ this method and investigate the warped product of two off-shell Klein--Gordon simple separable metrics. We prove that, although such a metric does not fit in the Klein--Gordon simple separable class, it leads again to a separable Klein--Gordon equation.

For this metric we solve the Einstein equations and prove that the component of the warped product under the warped factor must be the on-shell Klein--Gordon simple separable metric. Since this component must be Euclidean (temporal direction belongs to the other component), it is either an on-shell Kerr--NUT--(A)dS Euclidean instanton (a deformed, twisted sphere, cf.~\cite{KrtousEtal:2015}) or the on-shell Klein--Gordon simple separable Einstein--K\"ahler metric. The former solution has been recently obtained as a limit of Kerr--NUT--(A)dS spacetime where several rotations are sent to zero with properly scaled coordinates, mass, and NUT charges \cite{KrtousEtal:2015}. It also has been studied in the context of warped products of Kerr--NUT--(A)dS spaces \cite{KrtousKubiznakKolar:2015}. The latter case is, to our knowledge, a new solution of the Einstein equations in higher dimensions.

The paper is organized as follows. In Sec.\ \ref{sc:kgsepsol} we introduce the class of Klein--Gordon simple separable metrics and discuss some of their properties. In particular, we show that such a metric possesses  Killing vectors and rank-two Killing tensors, which enable us to solve the Klein--Gordon equation by separation of variables. Then, we solve the Einstein equations and, thus, obtain the on-shell Klein--Gordon simple separable spacetimes.

Section\ \ref{sc:warpkgsepsol} is devoted to the warped spacetimes. We briefly summarize the main properties of general warped spaces. Next, we consider a warped geometry of two Klein--Gordon simple separable metrics with a suitable chosen warped factor.
%Although the warped metric itself does not belong to this class, it inherits the key properties: it possesses the tower of Killing vectors and tensors and the Klein--Gordon equation is also separable.
Finally, adopting the warped metric ansatz, we find two families of solutions of the Einstein equations.

We conclude with a brief summary in Sec.\ \ref{sc:concl}. In Appendixes \ref{ap:ui} and \ref{ap:curv}, we gather some useful identities and expressions for the curvature of the studied metrics.

For simplicity, we restrict our discussion to even dimensions. Generalization to odd dimensions is straightforward, but one has to deal with additional ``odd'' terms in most expressions.

\section{Klein--Gordon simple separable spacetimes}\label{sc:kgsepsol}

\subsection{Off-shell metric}\label{ssc:kgsepsol:offmet}
Consider a metric of even dimension\footnote{%
In what follows, we do not assume an implicit sum over greek indices ${\mu,\nu,\dots}$ and latin indices ${i,j,\dots}$. Unless otherwise stated, indices have ranges ${\mu,\nu = 1,\dots,N}$ and ${i,j=0,\dots,N-1}$, respectively. We use here shortened notations ${\sum_\mu \equiv \sum_{\mu=1}^N}$, ${\sum_i \equiv \sum_{i=0}^{N-1}}$ and ${\prod_\mu \equiv \prod_{\mu=1}^N}$, ${\prod_i \equiv \prod_{i=0}^{N-1}}$.} ${D=2N}$,
\begin{equation}\label{eq:zmetric}
  \tb{g}=\sum_{\mu}\bigg[\frac{U_\mu}{X_\mu}\big(\tb{\dd}x_\mu\big)^{\!\bs{2}}
  +\frac{X_\mu}{U_\mu}\Big(\sum_{j}A_{\mu}^{(j)}\tb{\dd}\psi_j\Big)^{\!\bs{2}}\bigg]\;,
\end{equation}
where ${X_\mu=X_\mu(x_\mu)}$ are arbitrary functions of a single variable $x_\mu$. Functions $U_\mu$ and $A_{\mu}^{(i)}$ are defined by
\begin{equation}\label{eq:AUfunctions}
	U_\mu = \prod_{\mathclap{\substack{\nu\\ \nu\neq\mu}}} \big(Z_\nu-Z_\mu\big)\;,
	\quad
	A_{\mu}^{(i)} = \sum_{\mathclap{\substack{\nu_1,\dots,\nu_i \\ \nu_1<\dots<\nu_i \\ \nu_k\neq\mu}}} Z_{\nu_1} \dots Z_{\nu_i}\;.
\end{equation}
We set ${U_1=1}$ for ${N=1}$ and ${A_{\mu}^{(0)}=1}$. For further reference, we also define functions
\begin{equation}\label{eq:functionUA}
U=\prod_{\mathclap{\substack{\mu,\nu \\ \mu<\nu}}}\big(Z_\mu-Z_\nu\big)\;,
\quad
	A^{(i)} = \sum_{\mathclap{\substack{\nu_1,\dots,\nu_i \\ \nu_1<\dots<\nu_i}}} Z_{\nu_1} \dots Z_{\nu_i}\;,
\end{equation}
where we set ${U=1}$ for ${N=1}$ and ${A^{(0)}=1}$. Here, ${Z_\mu=Z_\mu(x_\mu)}$ are arbitrary functions of a single variable $x_\mu$.  In what follows, we also assume that all functions $Z_\mu$ are functionally independent, i.e., ${Z'_\mu\neq 0}$. However, some results can be generalized to the cases when ${Z_\mu=r_\mu}$, where $r_\mu$ are constants satisfying ${r_\mu\neq r_\nu}$, ${\mu\neq\nu}$.

Metric \eqref{eq:zmetric} generalizes the four-dimensional metric introduced by Carter in \cite{Carter:1968b,Carter:2009}, and since it leads also to the separable Klein--Gordon equation, we call it the Klein--Gordon simple separable metric.

It is also clearly motivated by the (off-shell) Kerr--NUT--(A)dS metric in even dimensions \cite{ChenLuPope:2006,ChenLuPope:2007}, which can be obtained by setting $Z_\mu=x_\mu^2$. We could also rewrite the metric in an alternative form,\
\begin{equation}\label{eq:XYmetric}
  \tb{g}=\sum_{\mu}\bigg[\frac{U_\mu}{Y_\mu}\big(\tb{\dd}x_\mu\big)^{\!\bs{2}}
  +\frac{X_\mu}{U_\mu}\Big(\sum_{j}A_{\mu}^{(j)}\tb{\dd}\psi_j\Big)^{\!\bs{2}}\bigg]\;,
\end{equation}
in which $Z_\mu$ is eliminated by introducing a new coordinate $x_\mu$ given exactly by relation $Z_\mu=x_\mu^2$; the freedom in $Z_\mu$ is then shifted to a suitable defined new metric function $Y_\mu=Y_\mu(x_\mu)$. In this form, the metric is clearly of type A of \cite{HouriEtal:2012}, which admits a generalized principal Killing--Yano tensor with torsion. The metric \eqref{eq:XYmetric} was also used in studying a higher-dimensional generalization of the Wahlquist metric \cite{HinoueEtal:2014}.\footnote{The off-shell Wahlquist metric can be regained from \eqref{eq:zmetric} by setting ${Z_\mu=\frac{1}{\beta^2}\sinh^2 \beta x_\mu}$.} In the following, we will use the original form \eqref{eq:zmetric} of the metric.

The determinant $\gden$ of the metric \eqref{eq:zmetric} in coordinates $x_\mu$, $\psi_k$ reads
\begin{equation}\label{eq:deter}
	\gden=U^2\;.
\end{equation}

We introduce the orthonormal frame ${\ev{\mu},\,\evh{\mu}}$, and dual covector frame ${\ef{\mu},\,\efh{\mu}}$,
\begin{align}\label{eq:ebase}
\begin{gathered}
	\ev{\mu} =\sqrt{\frac{X_\mu}{U_\mu}}\tvec{x_\mu}\;,\quad
	 \evh{\mu}=\sqrt{\frac{U_\mu}{X_\mu}}\sum_{k}\frac{\big(\!-\!Z_\mu\big)^{\!N-1-k}}{U_\mu}\tvec{\psi_k}\;,
	\\
	\ef{\mu}=\sqrt{\frac{U_\mu}{X_\mu}}\tb{\dd}x_\mu\;,\quad
	\efh{\mu}=\sqrt{\frac{X_\mu}{U_\mu}}\sum_{k}A_{\mu}^{(k)}\tb{\dd}\psi_k\;.
\end{gathered}\raisetag{5ex}
\end{align}

The Riemann tensor of metric \eqref{eq:zmetric} is given in \eqref{eq:zRiem} in Appendix. The Ricci tensor can be written in the form
\begin{equation}\label{eq:zRic}
\begin{split}
	\tb{\Ric} &= \sum_{\mu}T_\mu\ef{\mu}\ef{\mu} +\sum_{\mu}\bigg[T_\mu +\frac{1}{2}\sum_{\mathclap{\substack{\nu\\ \nu\neq\mu}}}\frac{S_{\nu,\mu}}{Z'_\mu}\frac{X_\mu}{U_\mu}	 \bigg]\efh{\mu}\efh{\mu}
	\\%%%%%%%%%%%%%%%%%%%%%%%%%%%%%%%%%%%%%%%%%%%%%%%%%%%%%%%%%%%%%
	&\feq +\sum_{\mathclap{\substack{\mu,\nu\\ \nu\neq\mu}}}\frac{1}{2}\frac{S_\mu-S_\nu}{Z_\mu-Z_\nu}\sqrt{\frac{X_\mu }{U_\mu}\frac{X_\nu }{U_\nu}}\efh{\mu}\efh{\nu}\;,
\end{split}\raisetag{7ex}
\end{equation}
where\footnote{%
Despite the zeros in the denominators, some of the following expressions can be also extended to the ${Z'_\nu= 0}$ case because ${S_{\mu,\nu}=-\frac{2Z'_\nu}{Z_\mu-Z_\nu}\bigg(Z''_\nu-\frac12\frac{{Z'_\mu}^2-{Z'_\nu}^2}{Z_\mu-Z_\nu}\bigg)}$.}
\begin{equation}\label{eq:rufunctions}
\begin{split}
	S_\mu &=Z''_\mu+\sum_{\mathclap{\substack{\kappa\\ \kappa\neq\mu}}}\frac{{Z'_\mu}^2-{Z'_\kappa}^2}{Z_\mu-Z_\kappa}\;,
		\\
		T_\mu &=-\frac{1}{2}\frac{X''_\mu}{U_\mu}-\frac{1}{2}\sum_{\mathclap{\substack{\nu\\ \nu\neq\mu}}}\frac{S_{\mu,\nu}}{Z'_\nu}\frac{X_\nu}{U_\nu}
	\\
	&\feq+\frac{1}{2}\sum_{\mathclap{\substack{\nu\\ \nu\neq\mu}}}\frac{1}{U_\mu}\bigg[\frac{Z'_\mu X'_\mu-Z''_\mu X_\mu}{Z_\mu-Z_\nu}+\frac{Z'_\nu X'_\nu-Z''_\nu X_\nu}{Z_\nu-Z_\mu}\bigg]\;.
\end{split}\raisetag{15ex}
\end{equation}
The scalar curvature reads
\begin{equation}\label{eq:zRsc}
	\mathcal{R}=-\sum_{\mu}\frac{X''_\mu}{U_\mu}\;.
\end{equation}

The metric \eqref{eq:zmetric} admits a generalized closed conformal Killing--Yano 2-form with torsion \cite{HouriEtal:2012}
\begin{equation}\label{eq:h2form}
	\tb{h} = \sum_\mu \sqrt{Z_\mu} \ef{\mu}\wedge\efh{\mu}\;,
\end{equation}
which guarantees the existence of hidden symmetries encoded by rank-two Killing tensors,
\begin{equation}\label{eq:killT}
\begin{split}
\tb[^{\ind{i}}]{k}
    &=\sum_{\mu}A_{\mu}^{(i)}\!\bigg[\frac{X_\mu}{U_\mu}\bigg(\!\tvec{x_\mu}\!\bigg)^{\!\!\bs{2}}
    {+}\frac{U_\mu}{X_\mu}\!\bigg(\!\!\sum_{k}\frac{\big({-}Z_\mu\big)^{\!N{-}1{-}k}}{U_\mu}\tvec{\psi_k}\!\bigg)^{\!\!\bs{2}}\bigg]
    \\
  &=\sum_{\mu}A_{\mu}^{(i)}\big(\ev{\mu}\ev{\mu}+\evh{\mu}\evh{\mu}\big)\;.\raisetag{5ex}
\end{split}
\end{equation}
The geometry \eqref{eq:zmetric} also possesses a set of explicit symmetries given by the Killing vectors,
\begin{equation}\label{eq:killV}
\tb[^{\ind{i}}]{l} =\tvec{\psi_i}=\sum_\mu \sqrt{\frac{X_\mu}{U_\mu}}A_{\mu}^{(i)}\evh{\mu}\;.
\end{equation}

Since the operators \eqref{eq:operKLwarped} constructed in the next subsection mutually commute, the corresponding phase-space observables $\tb[^{\ind{j}}]{k}[^{ab}]\tb{p}[_a]\tb{p}[_b]$, $\tb[^{\ind{k}}]{l}[^a]\tb{p}[_a]$ (quadratic and linear in momentum $\tb{p}$) must be in involution; for details see \cite{KolarKrtous:2015}. Thus, the geodesic motion is completely integrable and all Schouten--Nijenhuis brackets of $\tb[^{\ind{j}}]{k}$ and $\tb[^{\ind{k}}]{l}$ must vanish.

\subsection{Separability of the Klein--Gordon equation}\label{ssc:kgsepsol:sepkg}

We want to demonstrate that the geometry \eqref{eq:zmetric} enables us to solve the Klein--Gordon equation by a separation of variables. We will actually prove the separability for all operators based on the Killing tensors $\tb[^{\ind{j}}]{k}$ and Killing vectors $\tb[^{\ind{j}}]{l}$ introduced in  \eqref{eq:killT} and \eqref{eq:killV}, namely, for operators
\begin{equation}\label{eq:operKLwarped}
  {\op{K}}_j = -\tb{\nablai}[_a]\tb[^{\ind{j}}]{k}[^{ab}]\tb{\nablai}[_b]\;,\quad
  {\op{L}}_j =-\imag\,\tb[^{\ind{j}}]{l}[^a]\tb{\nablai}[_a]\;.
\end{equation}

Operator ${\op{K}}_j$ can be rewritten in coordinates using the fact that ${{\op{K}}_j= -\gden^{-\frac12}\tb{\pp}[_a]\gden^{\frac12}\tb[^{\ind{j}}]{k}[^{ab}]\tb{\pp}[_b]}$ and employing relation \eqref{eq:UdivUmu}. Here, $\tb{\pp}$ denotes the coordinate derivative with respect to the coordinates $x_\mu$, $\psi_k$. The operators ${\op{K}}_j$ and ${\op{L}}_j$ then read\footnote{%
In operator equations we use the convention that the round brackets around a derivative end the action of the derivative to the right, however, the square brackets do not. It means ${\big[\tb{\nablai}[_a]\tb{l}[^a]\big] = \tb{l}[^a]\tb{\nablai}[_a] + \big(\tb{\nablai}[_a]\tb{l}[^a]\big)}$. Applying the operator on a scalar ${\phi}$, we get ${\big[\tb{\nablai}[_a]\tb{l}[^a]\big]\phi = \big(\tb{\nablai}[_a](\tb{l}[^a]\phi)\big) = \tb{l}[^a]\big(\tb{\nablai}[_a]\phi\big) + \phi \big(\tb{\nablai}[_a]\tb{l}[^a]\big)}$.}
\begin{equation}\label{eq:operatorsKL}
\begin{split}
	{\op{K}}_j &=\sum_{\mu} \frac{A_{\mu}^{(j)}}{U_\mu}\Bigg[-\frac{\pp}{\pp x_\mu}X_\mu \frac{\pp}{\pp x_\mu}
	\\
	 &\feq+\frac{1}{X_\mu}\bigg[\sum_k\big(\!-\!Z_\mu\big)^{\!N-1-k}{\op{L}}_k\bigg]^2\Bigg]\;,
	\\
	{\op{L}}_k &=-\imag\frac{\pp}{\pp\psi_k}\;.
\end{split}
\end{equation}

Following the procedure in \cite{SergyeyevKrtous:2008}, we show that these operators mutually commute.
First, the commutators of ${\op{L}}_j$ among themselves and of ${\op{L}}_j$ and ${\op{K}}_k$ obviously vanish,
\begin{equation}
	\com{{\op{L}}_j}{{\op{L}}_k}=0\;,
	\quad
	\com{{\op{L}}_j}{{\op{K}}_k}=0\;,
\end{equation}
and it remains to prove that
\begin{equation}\label{eq:KKcom}
	\com{{\op{K}}_j}{{\op{K}}_k}=0\;.
\end{equation}
Operators ${\op{K}}_j$ can be written as
\begin{equation}\label{eq:KjMmu}
{\op{K}}_j =\sum_{\mu} \frac{A_{\mu}^{(j)}}{U_\mu}{\op{M}}_\mu\;,
\end{equation}
where
\begin{equation}
	{\op{M}}_\mu =-\frac{\pp}{\pp x_\mu}X_\mu \frac{\pp}{\pp x_\mu}
+\frac{1}{X_\mu}\bigg[\sum_k\big(\!-\!Z_\mu\big)^{\!N-1-k}{\op{L}}_k\bigg]^2\;.
\end{equation}
Operator ${\op{M}}_\mu$ depends only on a corresponding coordinate $x_\mu$ and contains a derivative $\frac{\pp}{\pp x_\mu}$. Due to this fact, these operators commute,
 \begin{equation}\label{eq:MmuMnu}
	\com{{\op{M}}_\mu}{{\op{M}}_\nu}=0\;.
\end{equation}
Substituting the inverse relation [cf. identities \eqref{eq:AUid}]
\begin{equation}
	{\op{M}}_\mu = \sum_{j}\big(\!-\!Z_\mu\big)^{\!N-1-j}{\op{K}}_j\;
\end{equation}
into \eqref{eq:MmuMnu}, we obtain
\begin{equation}
\sum_{j,k}\big(\!-\!Z_\mu\big)^{\!N-1-j}\big(\!-\!Z_\nu\big)^{\!N-1-k}\com{{\op{K}}_j}{{\op{K}}_k}=0\;,
\end{equation}
which leads to \eqref{eq:KKcom}.

Therefore, all operators \eqref{eq:operatorsKL} mutually commute and, thus, they have a common set of eigenfunctions, which satisfy
\begin{equation}
	{\op{K}}_j \phi =\Xi_j \phi\;,
	\quad
	{\op{L}}_j \phi =\Psi_j \phi\;.
\end{equation}
Here, $\Xi_j$ and $\Psi_j$ are corresponding eigenvalues which label the eigenfunctions.

By employing the identities \eqref{eq:AUid}, we find the solution in the separated form
\begin{equation}\label{eq:phiansatz}
	\phi = \prod_\mu R_\mu \prod_k\exp{\big(\imag\Psi_k\psi_k\big)}\;,
\end{equation}
where ${R_\mu=R_\mu(x_\mu)}$ are single-variable functions which satisfy ordinary differential equations,
\begin{equation}\label{eq:Rdifeq}
	\Big(X_\mu R'_{\mu} \Big)'+\bigg(\breve{\Xi}_\mu -\frac{\breve{\Psi}_\mu^2}{X_\mu}\bigg)R_\mu=0\;,
\end{equation}
where
\begin{equation}
	\breve{\Psi}_\mu=\sum_k\Psi_k\big(\!-\!Z_\mu\big)^{\!N-1-k}\;,
	\quad
	\breve{\Xi}_\mu=\sum_k\Xi_k\big(\!-\!Z_\mu\big)^{\!N-1-k}\;.
\end{equation}

\subsection{Solutions of the Einstein equations}\label{ssc:kgsepsol:solee}

In what follows, we find functions $Z_\mu$ and $X_\mu$ for which the metric \eqref{eq:zmetric} solves the vacuum Einstein equations with cosmological constant $\Lambda$, which in ${D=2N}$ can be written as
\begin{equation}
\tb{\Ric}=\frac{\Lambda}{N-1}\tb{g}\;, \quad N>1\;.
\end{equation}

We see from \eqref{eq:zRic} that we must require
\begin{equation}\label{eq:umuunu}
	S_\mu=S_\nu
\end{equation}
to satisfy the off-diagonal part of Einstein equations. In the four-dimensional case, ${N=2}$, this requirement says that ${Z''_\mu=2p}$, i.e.,
\begin{equation}\label{eq:zpoly}
	Z_\mu=p x_\mu^2 + q_\mu x_\mu+ r_\mu\;,
\end{equation}
with $p$, $q_\mu$, and $r_\mu$ being constants. Assumption ${Z'_\mu\neq 0}$ requires that ${p\neq 0}$ or ${p=0}$, ${q_\mu\neq 0}$.

If ${N>2}$, then \eqref{eq:umuunu} implies that functions $Z_\mu$ meet the condition
\begin{equation}\label{eq:condzK}
	\frac{{Z'_\mu}^2-{Z'_\nu}^2}{Z_\mu-Z_\nu}=K_\mu+K_\nu+s_{\mu\nu}
\end{equation}
for the single-variable functions ${K_\mu=K_\mu(x_\mu)}$ and constants ${s_{\mu\nu}=s_{\nu\mu}}$. Indeed, if we differentiate \eqref{eq:umuunu} with respect to $x_\kappa$, where ${\kappa\neq\mu,\nu}$, we find out that
\begin{equation}
	\bigg(\frac{{Z'_\mu}^2-{Z'_\kappa}^2}{Z_\mu-Z_\kappa}\bigg)_{\!\!,\kappa} = L_\kappa
\end{equation}
for the single-variable functions ${L_\kappa=L_\kappa(x_\kappa})$. Integrating this equation and using the symmetry of the left-hand side, we obtain the condition \eqref{eq:condzK}.

Next, we multiply the relation \eqref{eq:condzK} by ${Z_\mu-Z_\nu}$ and differentiate it with respect to both indices, which gives
\begin{equation}\label{eq:zK}
Z'_\mu K'_\nu=Z'_\nu K'_\mu\;.
\end{equation}
This equation is equivalent to
\begin{equation}\label{eq:Koverz}
	\frac{K'_{\mu}}{Z'_{\mu}}=q\;,
\end{equation}
with $q$ being a separation constant. By integrating \eqref{eq:Koverz}, we obtain
\begin{equation}\label{eq:KAzB}
	K_{\mu}=q Z_\mu+p_{\mu}\;,
\end{equation}
where $p_\mu$ are arbitrary constants. Substituting \eqref{eq:KAzB} into \eqref{eq:condzK} yields
\begin{equation}\label{eq:zeqABC}
	{Z'_{\mu}}^{2}-q Z_{\mu}^{2}-(p_{\mu}+p_{\nu}+s_{\mu\nu})Z_\mu=r_{\mu\nu}\;,\quad\mu\neq \nu\;,
\end{equation}
where ${r_{\mu\nu}=r_{\nu\mu}}$ are separation constants. Differentiating this relation we find that the expressions ${(p_{\mu}+p_{\nu}+s_{\mu\nu})}$ and $r_{\mu\nu}$ do not depend on indices $\mu$, $\nu$. We denote these constants by $4p$ and $r$, respectively, i.e.
\begin{equation}\label{eq:zeqABC2}
	{Z'_{\mu}}^{2}-q Z_{\mu}^{2}-4pZ_\mu=r\;.
\end{equation}
In order to ensure the compatibility with the original equation \eqref{eq:umuunu}, the constant $q$ must vanish. Thus, the relation \eqref{eq:zeqABC2}, finally, takes the form
\begin{equation}\label{eq:zeqABCfinal}
	{Z'_{\mu}}^{2}-4p Z_\mu=r\;.
\end{equation}

Equation \eqref{eq:zeqABCfinal} is satisfied if and only if $Z_\mu$ has the form \eqref{eq:zpoly}, where $r_{\mu}$ are arbitrary constants. However, unlike the ${N=2}$ case, the constants $q_\mu$ must now satisfy
\begin{equation}\label{eq:constrelation}
	q_\mu^2- 4p r_{\mu}=r,
\end{equation}
where ${p\neq 0}$ or ${p=0}$, ${r\neq 0}$.

Let us now turn our attention to the trace of the Einstein equations,
\begin{equation}
\mathcal{R}=\frac{2N\Lambda}{N-1}\;.
\end{equation}
Substituting \eqref{eq:zRsc} and applying lemma \ref{le:Fmueq}, we obtain
\begin{equation}\label{eq:difeqforXmu}
	 -X''_\mu=\frac{2N\Lambda}{N-1}\big(\!-\!Z_\mu\big)^{\!N-1}+\sum_{k=0}^{N-2}a_k\big(\!-\!Z_\mu\big)^{\!k}\;,
\end{equation}
with arbitrary constants $a_k$. Equation \eqref{eq:difeqforXmu} is a simple linear differential equation, where the right-hand side is a polynomial of degree at most $2N$ in $x_\mu$. It has a general solution,
\begin{equation}\label{eq:tracesol}
		X_{\mu} =\alpha_\mu x_\mu+\beta_\mu+\sum_{k=1}^{N+1}d_k\big(\!-\!Z_\mu\big)^{\!k}\;,
\end{equation}
where $\alpha_\mu$, $\beta_\mu$ are arbitrary coefficients of the homogeneous solution and $d_k$ correspond to the particular solution. These coefficients are determined by the right-hand side of \eqref{eq:difeqforXmu} or just replace the arbitrary constants $a_k$,
\begin{equation}\label{eq:constrel}
\begin{split}
	d_{N+1} &=
		\begin{cases}
			\mathrlap{0}\hphantom{\frac{1}{(N-1)(2N-1)}\frac{\Lambda}{p}}  &p\neq 0\\
			-\frac{2}{N^2-1}\frac{\Lambda}{r} &p=0,r\neq 0
		\end{cases}\;,
	\\
	d_{N} &=
		\begin{cases}
			\frac{1}{(N-1)(2N-1)}\frac{\Lambda}{p} &p\neq 0\\
			\const \in\mathbb{R} &p=0,r\neq 0
		\end{cases}\;,
	\\
	d_{N-1} &= \const \in\mathbb{R}\;,
	\\
	\vdots &
	\\
	d_1 &= \const \in\mathbb{R}\;.
\end{split}
\end{equation}
Substituting \eqref{eq:tracesol} into the full Einstein equations and employing \eqref{eq:AUid} and \eqref{eq:Uid}, we find
\begin{equation}\label{eq:constantsrelationaqbp}
	\alpha_\mu q_\mu-2\beta_\mu p=s\;,
\end{equation}
where $s$ is a constant.

Thus, the full Einstein equations are satisfied if and only if the functions $Z_\mu$ are given by \eqref{eq:zpoly}, \eqref{eq:constrelation}, and the functions $X_\mu$ by \eqref{eq:tracesol}--\eqref{eq:constantsrelationaqbp}.

These relations are also valid for ${N=2}$, because the condition \eqref{eq:zeqABCfinal} is met here anyway; however, here it is enforced by the diagonal part instead of the off-diagonal part of the Einstein equations.

For ${p\neq 0}$, we can introduce the new coordinates
\begin{equation}\label{eq:coortran1}
\begin{split}
	\check{x}_\mu &=\sqrt{p}\bigg(x_\mu+\frac{q_\mu}{2p}\bigg)\;,
	\\
	\check{\psi}_j &=\frac{1}{\sqrt{p}}\sum_{k=0}^{N-1-j}\frac{(j+k)!}{j!}\bigg(\!\!-\!\frac{r}{4p}\bigg)^{\!\! k}\psi_{j+k}\;.
\end{split}
\end{equation}
In these coordinates, the metric \eqref{eq:zmetric} takes the form
\begin{equation}\label{eq:zmetricnewcoorAnz}
  \tb{g}=\sum_{\mu}\bigg[\frac{\check{U}_\mu}{\check{X}_\mu}\big(\tb{\dd}\check{x}_\mu\big)^{\!\bs{2}}
  +\frac{\check{X}_\mu}{\check{U}_\mu}\Big(\sum_{j}\check{A}_{\mu}^{(j)}\tb{\dd}\check{\psi}_j\Big)^{\!\bs{2}}\bigg]\;,
\end{equation}
where
\begin{equation}\label{eq:solquadnew}
	\check{Z}_\mu =\check{x}_\mu^2\;,
	\quad
	\check{X}_\mu =\sum_{k=0}^{N}\check{c}_k\big(\!-\!\check{x}_\mu^2\big)^{\!k}+\check{b}_\mu \check{x}_\mu\;.
\end{equation}
Here, the coefficient of the highest power of $x_\mu$ is
\begin{equation}
	\check{c}_N=\frac{\Lambda}{(N-1)(2N-1)}\;,
\end{equation}
and the remaining coefficients $\check{c}_k$, $\check{b}_\mu$ are arbitrary constants replacing the $d$'s , $\alpha$'s, and $\beta$'s.
This is a standard form of the Kerr--NUT--(A)dS spacetime \cite{ChenLuPope:2006}, and the 2-form \eqref{eq:h2form} is a standard principal conformal Killing--Yano tensor. The metric is written in a Euclidean form, however, for a suitable choice of ranges of coordinates, it can be Wick-rotated to a Lorentzian signature (see \cite{ChenLuPope:2006,KrtousEtal:2015}).

For ${p=0}$ and ${r\neq 0}$, we define other coordinates\footnote{%
Symbol $\pm_\mu$ denotes the sign which depends on index $\mu$.}
\begin{equation}\label{eq:coortran2}
	\check{x}_\mu =\pm_\mu \sqrt{r}x_\mu+r_\mu\;,
	\quad
	\check{\psi}_j =\frac{1}{\sqrt{r}}\psi_j\;.
\end{equation}
In these coordinates, the metric \eqref{eq:zmetric} takes the form \eqref{eq:zmetricnewcoorAnz}, where
\begin{equation}\label{eq:sollinnew}
	\check{Z}_\mu =\check{x}_\mu\;,
	\quad
	\check{X}_\mu =\sum_{k=1}^{N+1}\check{c}_k\big(\!-\!\check{x}_\mu\big)^{\!k}+\check{b}_\mu\;.
\end{equation}
Here, the coefficient of the highest power of $x_\mu$ is
\begin{equation}
		\check{c}_{N+1}=-\frac{2\Lambda}{N^2-1}\;,
\end{equation}
and the remaining coefficients $\check{c}_k$ and $\check{b}_\mu$ are again arbitrary constants replacing the $d$'s , $\alpha$'s, and $\beta$'s.

It can be verified that the metric \eqref{eq:sollinnew} is K\"{a}hler and the \mbox{2-form}
\begin{equation}
	\tb{\Omega}=\sum_\mu \ef{\mu}\wedge\efh{\mu}\;
\end{equation}
is the corresponding K\"{a}hler 2-form. This Einstein--K\"{a}hler metric can be also obtained directly from the even-dimensional Kerr--NUT--(A)dS metric using a particular scaling limit \cite{Kubiznak:2009} and it is often used to construct a one-dimensional higher Sasakian metric.

Another approach of constructing these metrics is by employing an analog of the Euclidean BPS limit. Even and odd-dimensional Kerr--NUT--(A)dS spacetime then leads to Ricci-flat--K\"{a}hler and Einstein--Sasakian metrics, respectively, see \cite{ChenLuPope:2007,ChenLuPope:2006}.

Unfortunately, the metric \eqref{eq:sollinnew} has a Euclidean signature and cannot be simply Wick-rotated to a Lorentzian regime. This is why it was not much studied in ${D=4}$.

For further reference, we briefly discuss the ${N=1}$ case, when the metric \eqref{eq:zmetric} takes the form
\begin{equation}
  \tb{g}=\frac{1}{X_1}\big(\tb{\dd}x_1\big)^{\!\bs{2}}
  +X_1\big(\tb{\dd}\psi_0\big)^{\!\bs{2}}\;.
\end{equation}
The corresponding Ricci tensor and scalar curvature are
\begin{equation}\label{eq:2dmetcurv}
	\tb{\Ric} =-\frac{1}{2}X''_1\tb{g}\;,
	\quad
	\mathcal{R} =-X''_1\;.
\end{equation}
The closest two-dimensional analog of the Einstein equations reads ${\mathcal{R}-\Lambda=0}$, see \cite{Teitelboim:1983}.
This equation is met if and only if\vspace{-0.2cm}
\begin{equation}
	X_1 = \frac{\Lambda}{2}\big(\!-\!x_1^2\big)+b_1 x_1+c_0\;,
\end{equation}
\vspace{-0.2cm}
where $b_1$ and $c_0$ are arbitrary constants.

\section{Warped Klein--Gordon separable spacetimes}\label{sc:warpkgsepsol}

In the previous section we introduced the Klein--Gordon simple separable metric \eqref{eq:zmetric} and found two important solutions of the Einstein equations, \eqref{eq:solquadnew} and \eqref{eq:sollinnew}. Although the former is the Kerr--NUT--(A)dS spacetime, which has a clear physical significance, the latter cannot be even simply Wick-rotated to a Lorentzian signature. However, in higher dimensions, it could be used to construct physically interesting warped metrics. In addition, due to the simplicity of warped geometries, such metrics also could be Klein--Gordon separable.

\subsection{Warped geometry}\label{ssc:warpkgsepsol:warpgeom}

Consider the direct product ${M=\tilde{M}\times\bar{M}}$ of two manifolds with dimensions $\tilde{D}$ and $\bar{D}$, respectively. The warped space is the manifold $M$ of dimension ${D=\tilde{D}+\bar{D}}$ equipped with the metric
\begin{equation}\label{eq:warpedmetric}
	\tb{g} = \tb{\tilde{g}} + \tilde{w}^2\tb{\bar{g}}\;,
\end{equation}
where $\tb{\tilde{g}}$, $\tb{\bar{g}}$ are metrics on the manifolds $\tilde{M}$, $\bar{M}$, respectively, and $\tilde{w}^2$ is a warped factor. We assume that the tilded and barred objects depend only on a position in $\tilde{M}$ or $\bar{M}$, and they are also nontrivial only in tilded or barred directions.

We define torsion-free covariant derivatives $\tb{\nabla}$, $\tb{\tilde{\nabla}}$, and $\tb{\bar{\nabla}}$ satisfying ${\tb{\nabla}\tb{g}=0}$, ${\tb{\tilde{\nabla}}\tb{\tilde{g}}=0}$, and ${\tb{\bar{\nabla}}\tb{\bar{g}}=0}$, respectively.The Christofel coefficients with respect to the adjusted coordinates then satisfy the relation
\begin{equation}
	 \tb*{\Gamma}[^c_{ab}]=\tb*{\tilde{\Gamma}}[^c_{ab}]+\tb*{\bar{\Gamma}}[^c_{ab}]+\tb*{\Lambda}[^c_{ab}]\;,
\end{equation}
where
\begin{equation}
	 \tb*{\Lambda}[^c_{ab}]=2\tb{\tilde{\lambda}}[_{(a}]\tb*{\bar{\delta}}[^c_{b)}]-\tilde{w}^2\tb{\tilde{\lambda}}[^c]\tb*{\bar{g}}[_{ab}]\;
\end{equation}
with the 1-form $\tb{\tilde{\lambda}}$ being a logarithmic derivative of a warped factor,
\begin{equation}\label{eq:lambda}
	\tb{\tilde{\lambda}} =\tb{\dd}\ln{\tilde{w}}\;,\quad
	\tilde{\lambda}^2 =\tb{\tilde{g}}[^{ab}]\tb{\tilde{\lambda}}[_a]\tb{\tilde{\lambda}}[_b]\;.
\end{equation}

The Riemann tensor, Ricci tensor, and scalar curvature read
\begin{equation}\label{eq:warpcurv}
\begin{split}
	\tb{R}[_{ab}^c_d] &=\tb{\tilde{R}}[_{ab}^c_d]+\frac{2}{\tilde{w}}\tb{\tilde{H}}[_{d[a}]\tb*{\bar{\delta}}[^c_{b]}]-2\tilde{w}\,\tb{\tilde{g}}[^{ce}]\tb{\tilde{H}}[_{e[a}]\tb{\bar{g}}[_{b]d}]
	\\
	 &\feq+\tb{\bar{R}}[_{ab}^c_d]-2\tilde{w}^2\tilde{\lambda}^2\tb*{\bar{\delta}}[^c_{[a}]\tb{\bar{g}}[_{b]d}]\;,
	\\
	\tb{\Ric}[_{ab}]&=\tb{\tilde{\Ric}}[_{ab}]-\frac{\bar{D}}{\tilde{w}}\tb{\tilde{H}}[_{ab}]
	\\
	&\feq +\tb{\bar{\Ric}}[_{ab}]-\tilde{w}^2\bigg(\frac{\tilde{\mathcal{H}}}{\tilde{w}}+(\bar{D}-1)\tilde{\lambda}^2\bigg)\tb{\bar{g}}[_{ab}]\;,
	\\
	 \mathcal{R}&=\tilde{\mathcal{R}}+\frac{\bar{\mathcal{R}}}{\tilde{w}^2}-\frac{2\bar{D}}{\tilde{w}}\tilde{\mathcal{H}}-\bar{D}(\bar{D}-1)\tilde{\lambda}^2\;.
\end{split}
\end{equation}
Here, the Hessian tensor $\tb{\tilde{H}}$ and scalar $\tilde{\mathcal{H}}$ are
\begin{equation}\label{eq:hessian}
	\tb{\tilde{H}} =\tb{\tilde{\nabla}}\tb{\tilde{\nabla}}\tilde{w}\;,\quad
	\tilde{\mathcal{H}} = \tb{\tilde{g}}[^{ab}]\tb{\tilde{H}}[_{ab}]\;.
\end{equation}

The vacuum Einstein equations ${\tb{\Ric}=\frac{2\Lambda}{D-2}\tb{g}}$ split into two parts,
\begin{equation}\label{eq:EEtildebarA}
	\tb{\tilde{\Ric}} =\frac{\bar{D}}{\tilde{w}}\tb{\tilde{H}}+\frac{2\Lambda}{D-2}\tb{\tilde{g}}\;
\end{equation}
and
\begin{equation}\label{eq:EEtildebarB}
	\tb{\bar{\Ric}} =\tilde{w}^2\bigg(\frac{\tilde{\mathcal{H}}}{\tilde{w}}+(\bar{D}-1)\tilde{\lambda}^2+\frac{2\Lambda}{D-2}\bigg)\tb{\bar{g}}\;.
\end{equation}

However, the right-hand side of \eqref{eq:EEtildebarB} should not depend on the position in $\tilde{M}$, so the coefficient in front of $\tb{g}$ must be constant. We denote this constant by $\Upsilon$, i.e.,
\begin{equation}\label{eq:constantK}
	 \Upsilon=\tilde{w}^2\bigg(\frac{\tilde{\mathcal{H}}}{\tilde{w}}+(\bar{D}-1)\tilde{\lambda}^2+\frac{2\Lambda}{D-2}\bigg)\;.
\end{equation}

Assume that $\tb{\tilde{k}}$ and $\tb{\bar{k}}$ are rank-2 Killing tensors of the metrics $\tb{\tilde{g}}$ and $\tb{\bar{g}}$, respectively. Then it was shown in \cite{KrtousKubiznakKolar:2015} that
\begin{equation}\label{eq:KTwarped}
	\tb{\tilde{k}}+\frac{\tilde{A}}{\tilde{w}^2}\tb{\bar{g}^{-1}}
	\quad\text{and}\quad
	\tb{\bar{k}}
\end{equation}
are Killing tensors of the metric $\tb{g}$, provided that $\tilde{A}$ is a function on $\tilde{M}$ satisfying
\begin{equation}\label{eq:KTwarpedcond}
	\tb{\tilde{\nabla}}[^{a}]\!\tilde{A} = 2 \big(\tilde{A}\,\tb{\tilde{g}}[^{ab}]-\tb{\tilde{k}}[^{ab}]\big)\tb{\tilde{\lambda}}[_b]\;.
\end{equation}

\subsection{Off-shell metric}\label{ssc:warpkgsepsol:offmet}
Consider a warped metric \eqref{eq:warpedmetric}, where both $\tb{\tilde{g}}$ and $\tb{\bar{g}}$ are the off-shell metrics \eqref{eq:zmetric} of dimensions ${\tilde{D}=2\tilde{N}}$ and ${\bar{D}=2\bar{N}}$, respectively, and the warped function is given by\footnote{This form ensures the separability of the Klein--Gordon equation \eqref{eq:warpKGeq} below. It is probably not the most general ansatz with this property; the question of the most general warp factor admitting the separability remains open.}
\begin{equation}
	\tilde{w}^2=\tilde{A}^{(\tilde{N})}=\tilde{Z}_1\dots\tilde{Z}_{\tilde{N}}\;.
\end{equation}

The determinant $\gden$ of this metric in the coordinates $\tilde{x}_{\tilde{\mu}}$, $\tilde{\psi}_{\tilde{k}}$, $\bar{x}_{\bar{\mu}}$, $\bar{\psi}_{\bar{k}}$ reads
\begin{equation}\label{eq:deterwarp}
	\gden=\tilde{w}^{2\bar{N}}\tilde{U}^2\bar{U}^2\;.
\end{equation}
The 1-form \eqref{eq:lambda} is qiven by
\begin{equation}\label{eq:zLambda}
	\tb{\tilde{\lambda}} =\frac{1}{2}\sum_{\tilde{\mu}}\frac{\tilde{Z}'_{\tilde{\mu}}}{\tilde{Z}_{\tilde{\mu}}}\tb{\dd}\tilde{x}_{\tilde{\mu}}\;,\quad
	\tilde{\lambda}^2 =\frac{1}{4}\sum_{\tilde{\mu}}\frac{{{}\tilde{Z}'_{\tilde{\mu}}}^2}{\tilde{Z}_{\tilde{\mu}}^2}\frac{\tilde{X}_{\tilde{\mu}}}{\tilde{U}_{\tilde{\mu}}}\;.
\end{equation}
The Hessian tensor and scalar \eqref{eq:hessian} read
\begin{equation}\label{eq:zHessian}
\begin{split}
	\tb{\tilde{H}} &=\frac{1}{4}\sum_{\tilde{\mu},\tilde{\nu}} \frac{\tilde{w}}{\tilde{Z}'_{\tilde{\mu}}}\bigg[\bigg(\frac{{{}\tilde{Z}'_{\tilde{\nu}}}^2}{\tilde{Z}_{\tilde{\nu}}}\frac{\tilde{X}_{\tilde{\nu}}}{\tilde{U}_{\tilde{\nu}}}\bigg)_{\!\!\!,\tilde{\mu}}\eft{\tilde{\mu}}\eft{\tilde{\mu}} +\frac{{{}\tilde{Z}'_{\tilde{\nu}}}^2}{\tilde{Z}_{\tilde{\nu}}}\bigg(\frac{\tilde{X}_{\tilde{\nu}}}{\tilde{U}_{\tilde{\nu}}}\bigg)_{\!\!\!,\tilde{\mu}}\efht{\tilde{\mu}}\efht{\tilde{\mu}}\bigg]\;
	\\
	&\feq +\sum_{\mathclap{\substack{\tilde{\mu},\tilde{\nu}\\ \tilde{\nu}\neq\tilde{\mu}}}}\frac{\tilde{w}}{\tilde{Z}_{\tilde{\mu}}-\tilde{Z}_{\tilde{\nu}}}\bigg(\frac{{{}\tilde{Z}'_{\tilde{\nu}}}^2}{\tilde{Z}_{\tilde{\nu}}}-\frac{{{}\tilde{Z}'_{\tilde{\mu}}}^2}{\tilde{Z}_{\tilde{\mu}}}\bigg)\sqrt{\frac{\tilde{X}_{\tilde{\mu}}}{\tilde{U}_{\tilde{\mu}}}\frac{\tilde{X}_{\tilde{\nu}}}{\tilde{U}_{\tilde{\nu}}}}\efht{\tilde{\mu}}\efht{\tilde{\nu}}\;,
	\\
	\tilde{\mathcal{H}} &= \frac{1}{2}\sum_{\tilde{\mu}}\frac{\tilde{w}}{\sqrt{\tilde{Z}_{\tilde{\mu}}}}\Bigg(\frac{\tilde{Z}'_{\tilde{\mu}}}{\sqrt{\tilde{Z}_{\tilde{\mu}}}}\tilde{X}_{\tilde{\mu}}\Bigg)'\frac{1}{\tilde{U}_{\mu}}\;.
\\\end{split}\raisetag{6ex}
\end{equation}
The Riemann tensor, Ricci tensor, and scalar curvature can be then calculated directly by substituting \eqref{eq:zRiem}, \eqref{eq:zRic}, \eqref{eq:zRsc}, \eqref{eq:zHessian}, and \eqref{eq:zLambda} into the general expressions \eqref{eq:warpcurv}.

Employing \eqref{eq:KTwarped}, \eqref{eq:KTwarpedcond}, and \eqref{eq:zAAA}, it can be easily verified that
\begin{equation}\label{eq:killTwarped}
	\tb[^{\ind{\tilde{i}}}]{k} = \tb[^{\ind{\tilde{i}}}]{\tilde{k}}+\frac{\tilde{A}^{(\tilde{i})}}{\tilde{w}^2}\tb{\bar{g}^{-1}}\;,
		\quad
	\tb[^{\ind{\tilde{N}+\bar{i}}}]{k} = \tb[^{\ind{\bar{i}}}]{\bar{k}}\;
\end{equation}
are rank-2 Killing tensors of the warped metric. Because the warped function $\tilde{w}$ does not depend on the coordinates $\tilde{\psi}_{\tilde{j}}$, the Killing vectors remain unchanged, i.e.,
\begin{equation}\label{eq:killVwarped}
	\tb[^{\ind{\tilde{i}}}]{l} = \tb[^{\ind{\tilde{i}}}]{\tilde{l}}\;,
		\quad
	\tb[^{\ind{\tilde{N}+\bar{i}}}]{l} = \tb[^{\ind{\bar{i}}}]{\bar{l}}\;.
\end{equation}
Again, since we will show that all operators \eqref{eq:operKL} constructed below mutually commute, the corresponding phase-space observables $\tb[^{\ind{j}}]{k}[^{ab}]\tb{p}[_a]\tb{p}[_b]$, $\tb[^{\ind{k}}]{l}[^a]\tb{p}[_a]$ must be in involution. Thus, the geodesic motion is completely integrable and all Schouten--Nijenhuis brackets of $\tb[^{\ind{j}}]{k}$ and $\tb[^{\ind{k}}]{l}$ must vanish.

\subsection{Separability of the Klein--Gordon equation}\label{ssc:warpkgsepsol:sepkg}
We again define operators
\begin{equation}\label{eq:operKL}
  {\op{K}}_j = -\tb{\nablai}[_a]\tb[^{\ind{j}}]{k}[^{ab}]\tb{\nablai}[_b]\;,
  \quad
  {\op{L}}_j =-\imag\tb[^{\ind{j}}]{l}[^a]\tb{\nablai}[_a]\;,
\end{equation}
based on Killing tensors $\tb[^{\ind{j}}]{k}$ \eqref{eq:killTwarped} and Killing vectors $\tb[^{\ind{j}}]{l}$ \eqref{eq:killVwarped}. Operators \eqref{eq:operKL} can be also expressed in coordinates $\tilde{x}_{\tilde{\mu}}$, $\tilde{\psi}_{\tilde{k}}$, $\bar{x}_{\bar{\mu}}$, $\bar{\psi}_{\bar{k}}$ using ${{\op{K}}_j= -\gden^{-\frac12}\tb{\pp}[_a]\gden^{\frac12}\tb[^{\ind{j}}]{k}[^{ab}]\tb{\pp}[_b]}$ and employing relation \eqref{eq:UdivUmu},
\begin{equation}\label{eq:KKLLwarp}
\begin{split}
	{\op{K}}_{\tilde{j}} &= \sum_{\tilde{\mu}} \frac{\tilde{A}_{\tilde{\mu}}^{(\tilde{j})}}{\tilde{U}_{\tilde{\mu}}}\Bigg[-\frac{1}{\tilde{Z}_{\tilde{\mu}}^{\frac{\bar{N}}{2}}}\frac{\pp}{\pp \tilde{x}_{\tilde{\mu}}}\tilde{Z}_{\tilde{\mu}}^{\frac{\bar{N}}{2}}\tilde{X}_{\tilde{\mu}} \frac{\pp}{\pp \tilde{x}_{\tilde{\mu}}}
	\\
	 &\feq+\frac{1}{\tilde{X}_{\tilde{\mu}}}\bigg[\sum_{\tilde{k}}\big(\!-\!\tilde{Z}_{\tilde{\mu}}\big)^{\!\tilde{N}-1-\tilde{k}}{\op{L}}_{\tilde{k}}\bigg]^2\Bigg]+\frac{\tilde{A}^{(\tilde{j})}}{\tilde{A}^{(\tilde{N})}}{\op{K}}_{\tilde{N}}\;,
	 \\
	{\op{K}}_{\tilde{N}+\bar{j}} &=\sum_{\bar{\mu}} \frac{\bar{A}_{\bar{\mu}}^{(\bar{j})}}{\bar{U}_{\bar{\mu}}}\Bigg[-\frac{\pp}{\pp \bar{x}_{\bar{\mu}}}\bar{X}_{\bar{\mu}} \frac{\pp}{\pp \bar{x}_{\tilde{\mu}}}
	\\
	 &\feq+\frac{1}{\bar{X}_{\bar{\mu}}}\bigg[\sum_{\bar{k}}\big(\!-\!\bar{Z}_{\bar{\mu}}\big)^{\!\bar{N}-1-\bar{k}}{\op{L}}_{\tilde{N}+\bar{k}}\bigg]^2\Bigg]\;,
	\\
	{\op{L}}_{\tilde{k}} &=-\imag\frac{\pp}{\pp\tilde{\psi}_{\tilde{k}}}\;,
	\quad
	{\op{L}}_{\tilde{N}+\bar{k}} =-\imag\frac{\pp}{\pp\bar{\psi}_{\bar{k}}}\;.
\end{split}\raisetag{20ex}
\end{equation}

It can be shown that these operators mutually commute. Indeed, commutators  of ${\op{L}}_j$ among themselves and commutators of ${\op{L}}_j$ and ${\op{K}}_k$ vanish,
\begin{equation}
	\com{{\op{L}}_j}{{\op{L}}_k}=0\;,
	\quad
	\com{{\op{L}}_j}{{\op{K}}_k}=0\;.
\end{equation}
Operators ${\op{K}}_j$ have the form
\begin{equation}
\begin{split}
	{\op{K}}_{\tilde{j}} = \sum_{\tilde{\mu}} \frac{\tilde{A}_{\tilde{\mu}}^{(\tilde{j})}}{\tilde{U}_{\tilde{\mu}}}{\tilde{\op{M}}}_{\tilde{\mu}}+\frac{\tilde{A}^{(\tilde{j})}}{\tilde{A}^{(\tilde{N})}}{\op{K}}_{\tilde{N}}\;,
	 \quad
	{\op{K}}_{\tilde{N}+\bar{j}} =\sum_{\bar{\mu}} \frac{\bar{A}_{\bar{\mu}}^{(\bar{j})}}{\bar{U}_{\bar{\mu}}}{\bar{\op{M}}}_{\bar{\mu}}\;,
\end{split}
\end{equation}
where
\begin{equation}
\begin{split}
	{\tilde{\op{M}}}_{\tilde{\mu}} &= -\frac{1}{\tilde{Z}_{\tilde{\mu}}^{\frac{\bar{N}}{2}}}\frac{\pp}{\pp \tilde{x}_{\tilde{\mu}}}\tilde{Z}_{\tilde{\mu}}^{\frac{\bar{N}}{2}}\tilde{X}_{\tilde{\mu}} \frac{\pp}{\pp \tilde{x}_{\tilde{\mu}}}
		\\
		 &\feq+\frac{1}{\tilde{X}_{\tilde{\mu}}}\bigg[\sum_{\tilde{k}}\big(\!-\!\tilde{Z}_{\tilde{\mu}}\big)^{\!\tilde{N}-1-\tilde{k}}{\op{L}}_{\tilde{k}}\bigg]^2\;,
	\\
	{\bar{\op{M}}}_{\bar{\mu}} &=-\frac{\pp}{\pp \bar{x}_{\bar{\mu}}}\bar{X}_{\bar{\mu}} \frac{\pp}{\pp \bar{x}_{\bar{\mu}}} +\frac{1}{\bar{X}_{\bar{\mu}}}\bigg[\sum_{\bar{k}}\big(\!-\!\bar{Z}_{\bar{\mu}}\big)^{\!\bar{N}-1-\bar{k}}{\op{L}}_{\tilde{N}+\bar{k}}\bigg]^2\;.
\end{split}\raisetag{14ex}
\end{equation}
Commutators of ${\tilde{\op{M}}}_{\tilde{\mu}}$ and ${\bar{\op{M}}}_{\bar{\mu}}$ among themselves vanish,
\begin{equation}
	\com{{\tilde{\op{M}}}_{\tilde{\mu}}}{{\tilde{\op{M}}}_{\tilde{\nu}}}=0\;,
	\quad
	\com{{\bar{\op{M}}}_{\bar{\mu}}}{{\bar{\op{M}}}_{\bar{\nu}}}=0\;,	
\end{equation}
which, much like in Sec. \ref{ssc:kgsepsol:sepkg}, implies
\begin{equation}\label{eq:KtilKtilKbarKbar}
	\com{{\op{K}}_{\tilde{j}}}{{\op{K}}_{\tilde{k}}}=0\;,
	\quad
	\com{{\op{K}}_{\tilde{N}+\bar{j}}}{{\op{K}}_{\tilde{N}+\bar{k}}}=0\;.
\end{equation}
In the first equation we have also used the relation
\begin{equation}
	\comLR{\sum_{\tilde{\mu}} \frac{\tilde{A}_{\tilde{\mu}}^{(\tilde{j})}}{\tilde{U}_{\tilde{\mu}}}{\tilde{\op{M}}}_{\tilde{\mu}}}{\frac{\tilde{A}^{(\tilde{k})}}{\tilde{A}^{(\tilde{N})}}{\op{K}}_{\tilde{N}}}=\comLR{\sum_{\tilde{\mu}} \frac{\tilde{A}_{\tilde{\mu}}^{(\tilde{k})}}{\tilde{U}_{\tilde{\mu}}}{\tilde{\op{M}}}_{\tilde{\mu}}}{\frac{\tilde{A}^{(\tilde{j})}}{\tilde{A}^{(\tilde{N})}}{\op{K}}_{\tilde{N}}}\;,
\end{equation}
which is a consequence of identity \eqref{eq:DAiAn}. Finally, using the second equation in \eqref{eq:KtilKtilKbarKbar}, we can also prove the relation
\begin{equation}
	\com{{\op{K}}_{\tilde{j}}}{{\op{K}}_{\tilde{N}+\bar{k}}}=0\;.
\end{equation}

We have thus proved that operators \eqref{eq:KKLLwarp} mutually commute and, therefore, they have a common set of eigenfunctions; i.e., they satisfy
\begin{equation}\label{eq:warpKGeq}
	{\op{K}}_j \phi =\Xi_j \phi\;,
	\quad
	{\op{L}}_j \phi =\Psi_j \phi\;,
\end{equation}
where $\Xi_j$ and $\Psi_j$ are corresponding eigenvalues.

Employing identities \eqref{eq:AUid} and \eqref{eq:zAUid}, we find the solution in the separated form,
\begin{equation}\label{eq:phiansatzwarp}
	\phi = \prod_{\tilde{\mu}} \tilde{R}_{\tilde{\mu}} \prod_{\bar{\mu}} \bar{R}_{\bar{\mu}} \prod_{\tilde{k}}\exp{\big(\imag\Psi_{\tilde{k}}\tilde{\psi}_{\tilde{k}}\big)}\prod_{\bar{k}}\exp{\big(\imag\Psi_{\tilde{N}+\bar{k}}\bar{\psi}_{\bar{k}}\big)}\;,
\end{equation}
where ${\tilde{R}_{\tilde{\mu}}=\tilde{R}_{\tilde{\mu}}(\tilde{x}_{\tilde{\mu}})}$ and ${\bar{R}_{\bar{\mu}}=\bar{R}_{\bar{\mu}}(\bar{x}_{\bar{\mu}})}$ are single-variable functions which satisfy ordinary differential equations,
\begin{equation}\label{eq:Rdifeqwarp}
\begin{gathered}
	\Big(\tilde{Z}_{\tilde{\mu}}^{\frac{\bar{N}}{2}} \tilde{X}_{\tilde{\mu}} \tilde{R}'_{\tilde{\mu}} \Big)'+\bigg(\breve{\tilde{\Xi}}_{\tilde{\mu}} -\frac{\breve{\tilde{\Psi}}_{\tilde{\mu}}^2}{\tilde{X}_{\tilde{\mu}}}+\frac{\Xi_{\tilde{N}}}{\tilde{Z}_{\tilde{\mu}}}\bigg)\tilde{Z}_{\tilde{\mu}}^{\frac{\bar{N}}{2}}\tilde{R}_{\tilde{\mu}}=0\;,
	\\
	\Big(\bar{X}_{\bar{\mu}} \bar{R}'_{\bar{\mu}} \Big)'+\bigg(\breve{\bar{\Xi}}_{\bar{\mu}} -\frac{\breve{\bar{\Psi}}_{\bar{\mu}}^2}{\bar{X}_{\bar{\mu}}}\bigg)\bar{R}_{\bar{\mu}}=0\;,
\end{gathered}	
\end{equation}
where
\begin{equation}
\begin{split}
		\breve{\tilde{\Xi}}_{\tilde{\mu}} &=\sum_{\tilde{k}}\Xi_{\tilde{k}}\big(\!-\!\tilde{Z}_{\tilde{\mu}}\big)^{\!\tilde{N}-1-\tilde{k}}\;,
	\\
	\breve{\tilde{\Psi}}_{\tilde{\mu}} &=\sum_{\tilde{k}}\Psi_{\tilde{k}}\big(\!-\!\tilde{Z}_{\tilde{\mu}}\big)^{\!\tilde{N}-1-\tilde{k}}\;,
	\\
	\breve{\bar{\Xi}}_{\bar{\mu}} &=\sum_{\bar{k}}\Xi_{\tilde{N}+\bar{k}}\big(\!-\!\bar{Z}_{\bar{\mu}}\big)^{\!\bar{N}-1-\bar{k}}\;,
	\\
	\breve{\bar{\Psi}}_{\bar{\mu}} &=\sum_{\bar{k}}\Psi_{\tilde{N}+\bar{k}}\big(\!-\!\bar{Z}_{\bar{\mu}}\big)^{\!\bar{N}-1-\bar{k}}\;.
\end{split}
\end{equation}

\subsection{Solutions of the Einstein equations}\label{ssc:warpkgsepsol:solee}
In the following, we find functions $\tilde{Z}_{\tilde{\mu}}$, $\tilde{X}_{\tilde{\mu}}$, $\bar{Z}_{\bar{\mu}}$, and $\bar{X}_{\bar{\mu}}$ for which the warped metric solves the vacuum Einstein equations.

We see from the off-diagonal part of \eqref{eq:EEtildebarA} that
\begin{equation} \label{eq:warpedumuunu}
	 \tilde{S}_{\tilde{\mu}}+\bar{N}\frac{{{}\tilde{Z}'_{\tilde{\mu}}}^2}{\tilde{Z}_{\tilde{\mu}}}=\tilde{S}_{\tilde{\nu}}+\bar{N}\frac{{{}\tilde{Z}'_{\tilde{\nu}}}^2}{\tilde{Z}_{\tilde{\nu}}}\;.
\end{equation}
In the ${\tilde{N}=2}$ case, this condition says that
\begin{equation}\label{eq:warpedumuunun2}
	\tilde{Z}''_{\tilde{\mu}} + \bar{N}\frac{{{}\tilde{Z}'_{\tilde{\mu}}}^2}{\tilde{Z}_{\tilde{\mu}}} = 2\tilde{p}(2\bar{N}+1)\;,
\end{equation}
where $\tilde{p}$ is an arbitrary constant.

If ${\tilde{N}>2}$, then \eqref{eq:warpedumuunu} implies
\begin{equation}\label{eq:zeqABC2tilde}
	{{}\tilde{Z}'_{\tilde{\mu}}}^{2}-\tilde{q} \tilde{Z}_{\tilde{\mu}}^{2}-4\tilde{p}\tilde{Z}_{\tilde{\mu}}=\tilde{r}
\end{equation}
similar to the relation between \eqref{eq:umuunu} and \eqref{eq:zeqABC2}. However, in order to satisfy \eqref{eq:warpedumuunu}, the constants $\tilde{q}$ and $\tilde{r}$ must vanish. Thus, the relation \eqref{eq:zeqABC2tilde} takes the form
\begin{equation}\label{eq:zeqABFfinaltilde}
	{{}\tilde{Z}'_{\tilde{\mu}}}^{2}-4\tilde{p}\tilde{Z}_{\tilde{\mu}}=0\;,
\end{equation}
with $\tilde{p}$ being a nonzero constant. Equation \eqref{eq:zeqABFfinaltilde} is satisfied if and only if
\begin{equation}\label{eq:zpolytilde}
	\tilde{Z}_{\tilde{\mu}}=\tilde{p} \tilde{x}_{\tilde{\mu}}^2 + \tilde{q}_{\tilde{\mu}} \tilde{x}_{\tilde{\mu}}+ \tilde{r}_{\tilde{\mu}} \;,
\end{equation}
where
\begin{equation}\label{eq:constrelationtilde}
	\tilde{q}_{\tilde{\mu}}^2-4\tilde{p} \tilde{r}_{\tilde{\mu}}=0\;.
\end{equation}

In the ${\tilde{N}=1,2}$ cases, the form of \eqref{eq:zpolytilde} with the condition \eqref{eq:constrelationtilde} is enforced by the diagonal part of \eqref{eq:EEtildebarA}. In particular, in the ${\tilde{N}=1}$ case, it simply implies condition \eqref{eq:zeqABFfinaltilde}. For ${\tilde{N}=2}$, it gives
\begin{equation}
	 \bigg(\bar{N}\frac{{{}\tilde{Z}'_{\tilde{\mu}}}^2}{\tilde{Z}_{\tilde{\mu}}}+\frac{{{}\tilde{Z}'_{\tilde{\mu}}}^2-{{}\tilde{Z}'_{\tilde{\nu}}}^2}{\tilde{Z}_{\tilde{\mu}}-\tilde{Z}_{\tilde{\nu}}}\bigg)_{\!\!\!,\tilde{\mu}}=0\;.
\end{equation}
Integrating this equation, multiplying by ${\tilde{Z}_{\tilde{\mu}}-\tilde{Z}_{\tilde{\nu}}}$, and differentiating with respect to $\tilde{x}_{\tilde{\mu}}$ and $\tilde{x}_{\tilde{\nu}}$, we find that
\begin{equation}
	 \frac{1}{\tilde{Z}'_{\tilde{\mu}}}\bigg(\frac{{{}\tilde{Z}'_{\tilde{\mu}}}^2}{\tilde{Z}_{\tilde{\mu}}}\bigg)'= \tilde{t}_{\tilde{\mu}}\;,
\end{equation}
where $\tilde{t}_{\tilde{\mu}}$ is a separation constant. This equation leads to
\begin{equation}\label{eq:zpsqzpp}
	{{}\tilde{Z}'_{\tilde{\mu}}}^2=\tilde{t}_{\tilde{\mu}} \tilde{Z}_{\tilde{\mu}}^2 +\tilde{u}_{\tilde{\mu}} \tilde{Z}_{\tilde{\mu}}\;,
	\quad
	\tilde{Z}''_{\tilde{\mu}}=\tilde{t}_{\tilde{\mu}} \tilde{Z}_{\tilde{\mu}} +\frac12\tilde{u}_{\tilde{\mu}}\;,
\end{equation}
with $\tilde{u}_{\tilde{\mu}}$ being an arbitrary constant. Substituting \eqref{eq:zpsqzpp} into \eqref{eq:warpedumuunun2}, we determine the constants ${\tilde{t}_{\tilde{\mu}}=0}$ and ${\tilde{u}_{\tilde{\mu}}=4\tilde{p}}$ which yields \eqref{eq:zeqABFfinaltilde}.

Since $\tilde{Z}_{\tilde{\mu}}$ is a quadratic function \eqref{eq:zpolytilde} with ${\tilde{p}\neq 0}$, we can employ the coordinate freedom to transform the tilded coordinates using transformation \eqref{eq:coortran1} and drop the caron sign. This transformation preserves the form of the orthonormal frame, and it effectively sets the functions $\tilde{Z}_{\tilde{\mu}}$ to $\tilde{x}_{\tilde{\mu}}^2$. Equation \eqref{eq:constantK} can then be rewritten as
\begin{equation}
	 \sum_{\tilde{\mu}}\frac{1}{\tilde{U}_{\tilde{\mu}}}\bigg[\frac{\tilde{X}'_{\tilde{\mu}}}{\tilde{x}_{\tilde{\mu}}}+(2\bar{N}-1)\frac{\tilde{X}_{\tilde{\mu}}}{\tilde{x}_{\tilde{\mu}}^2}-\frac{\Upsilon}{\tilde{x}_{\tilde{\mu}}^2}+\frac{\Lambda}{N-1}\big(\!-\!\tilde{x}_{\tilde{\mu}}^2\big)^{\!\tilde{N}-1}\bigg]=0\;.
\end{equation}
Substituting \eqref{eq:zRsc} and applying lemma \ref{le:Fmueq}, we obtain
\begin{equation}\label{eq:eqfXkk}
\tilde{x}_{\tilde{\mu}}\tilde{X}'_{\tilde{\mu}}+(2\bar{N}-1)\tilde{X}_{\tilde{\mu}}=\Upsilon+\sum_{\tilde{k}=0}^{\tilde{n}-2} \tilde{a}_{\tilde{k}} \tilde{x}_{\tilde{\mu}}^{2(\tilde{k}+1)}+\frac{\Lambda}{N-1}\big(\!-\!\tilde{x}_{\tilde{\mu}}^2\big)^{\!\tilde{N}}\;.
\end{equation}

Equation \eqref{eq:eqfXkk} is a simple linear differential equation, where the right-hand side is a polynomial of degree $2\tilde{N}$ in $\tilde{x}_{\tilde{\mu}}$. It has a general solution,
\begin{equation}\label{eq:Xsoltilde}
	 \tilde{X}_{\tilde{\mu}}=\sum_{\tilde{k}=0}^{\tilde{N}}\tilde{c}_{\tilde{k}}\big(\!-\!\tilde{x}_\mu^2\big)^{\!\tilde{k}}+\frac{\tilde{b}_{\tilde{\mu}}}{\tilde{x}_{\tilde{\mu}}^{2\bar{N}-1}}\;,
\end{equation}
where
\begin{equation}\label{eq:Xsoltildeconst}
	\tilde{c}_{\tilde{N}}=\frac{\Lambda}{(2N-1)(N-1)}\;,
	\quad
\tilde{c}_{0}=\frac{\Upsilon}{2\bar{N}-1}\;,
\end{equation}
and coefficients $\tilde{b}_{\tilde{\mu}}$,  $\tilde{c}_{\tilde{k}}$, ${\tilde{k}\in 1,\dots,\tilde{N}-1}$ are arbitrary constants. We thus found metric functions for which the proportionality factor \eqref{eq:constantK} is a constant. Surprisingly, employing identities \eqref{eq:AUid}, the tilded part \eqref{eq:EEtildebarA} of the Einstein equations is already satisfied.

The barred part \eqref{eq:EEtildebarB} of the Einstein equations leads to the solutions \eqref{eq:solquadnew} or \eqref{eq:sollinnew} with the cosmological constant ${\bar\Lambda=(\bar{N}-1)\Upsilon}$, i.e.,
\begin{equation}\label{eq:Xsolbar1}
\bar{X}_{\bar{\mu}} =\sum_{\bar{k}=0}^{\bar{N}}\bar{c}_{\bar{k}}\big(\!-\!\bar{x}_{\bar{\mu}}^2\big)^{\!\bar{k}}+\bar{b}_{\bar{\mu}} \bar{x}_{\bar{\mu}}\;,
\quad
\bar{c}_{\bar{N}}=\frac{\Upsilon}{2\bar{N}-1}\;,
\end{equation}
where $\bar{b}_{\bar{\mu}}$, $\bar{c}_{k}$, ${\bar{k}\in 0,\dots,\bar{N}-1}$ are arbitrary constants,
or
\begin{equation}\label{eq:Xsolbar2}
\bar{X}_{\bar{\mu}} =\sum_{\bar{k}=1}^{\bar{N}+1}\bar{c}_{\bar{k}}\big(\!-\!\bar{x}_{\bar{\mu}}\big)^{\!\bar{k}}+\bar{b}_{\bar{\mu}} \;,
\quad
\bar{c}_{\bar{N}+1}=-\frac{2\Upsilon}{\bar{N}+1}\;,
\end{equation}
where $\bar{b}_{\bar{\mu}}$, $\bar{c}_{k}$, ${\bar{k}\in 1,\dots,\bar{N}}$ are arbitrary constants.

The warped solution with barred functions \eqref{eq:Xsolbar1} have been recently obtained in \cite{KrtousEtal:2015} and, as discussed there, it also belongs to class of generalized Kerr--NUT--(A)dS spacetimes \cite{HouriEtal:2008b,HouriEtal:2009}. The warped metric with barred functions \eqref{eq:Xsolbar2} constitutes new solutions of the Einstein equations.

\section{Conclusions}\label{sc:concl}

In this paper we have generalized Carter's ansatz for the metric leading to a separable Klein--Gordon equation to higher dimensions and thus defined the class of Klein--Gordon simple separable metrics. For these metrics, we have solved the Klein--Gordon equation by separation of variables, employing the symmetries encoded in Killing vectors and rank-two Killing tensors. The class of such metrics enlarges previously studied off-shell Kerr-NUT--(A)dS spaces. Solving the Einstein equations in this class, we have regained the on-shell Kerr--NUT--(A)dS spacetimes and the Euclidean Einstein--K\"ahler metrics which, as demonstrated in \cite{Kubiznak:2009,ChenLuPope:2007,ChenLuPope:2006}, can be obtained as a particular limit of the Kerr--NUT--(A)dS spaces as well.

Next, we have studied the warped spaces of the Klein--Gordon simple separable geometries. In this way we have constructed a more general metric for  which it is also possible to solve the Klein--Gordon equation by separation of variables. Although not all of these geometries possesses the principal closed conformal Killing--Yano form, they are equipped with the tower of rank-two Killing tensors and Killing vectors. These symmetry objects allow us to define phase-space variables, which are in involutions and guarantee that geodesic motion is completely integrable. They also define operators which commute with each other and the common set of eigenfunctions has been found by separation of variables.

Finally, we have solved the Einstein equations for the warped metric. We have recovered the spaces which can be obtained as a limit of vanishing rotations from the Kerr--NUT--(A)dS metric \cite{KrtousEtal:2015}. Beside them we have found spacetimes for which the metric block under the warp factor is the Klein--Gordon simple separable Einstein--K\"ahler metric. These spacetimes are new solutions in higher dimensions.

\section*{Acknowledgments}
This work was supported by the project of excellence of the Czech Science Foundation No.~\mbox{14-37086G}. I.K. was also supported by the Charles University Grant No.~\mbox{SVV-260211}. The authors would like to thank Filip Hejda for valuable discussions on Carter's paper% and Martin \v{Z}ofka for discussing the manuscript.

%%%%%%%%%%%%%%%%%%%%%%%%%%%%%%%%%%%%%%%%%%%%%%%%%%%%%%%%%%%%%%%%%%%%%%%%%%%%%%%%%%%%%%%%%%%%%%%%%%%%%%%%%%%%
%% APPENDIX

\appendix
\section{Useful identities}\label{ap:ui}
We list here some important identities for functions $A_{\mu}^{(i)}$, $U_\mu$ which emerge in the definition of metric \eqref{eq:zmetric}:
\begin{gather}
\label{eq:AUid}
\begin{gathered}
	\sum_\mu \frac{A_{\mu}^{(i)}}{U_\mu} \big(\!-\!Z_\mu\big)^{\!N-1-j} =\delta_{ij}\;, \\
	\sum_i  \frac{A_{\mu}^{(i)}}{U_\nu}\big(\!-\!Z_\nu\big)^{\!N-1-i} =\delta_{\mu\nu}\;,
\end{gathered}
\\
\label{eq:Uid}
 	\frac{1}{U_\nu}\sum_{\substack{\mu\\ \mu\neq\nu}}\frac{1}{Z_\mu-Z_\nu}=\sum_{\substack{\mu\\ \mu\neq\nu}}\frac{1}{U_\mu}\frac{1}{Z_\nu-Z_\mu}\;,
\\
\label{eq:zAAA}
	A^{(i)}=A_{\mu}^{(i)}+Z_\mu A_{\mu}^{(i-1)}\;,	
\\
\label{eq:zAUid}
	\sum_\mu\frac{A_{\mu}^{(j)}}{Z_\mu U_\mu}=\frac{A^{(j)}}{A^{(N)}}\;, 	
\\
\label{eq:Uder}
 	U_{\mu,\nu} =\delta_{\mu\nu}\sum_{\substack{\alpha\\ \alpha\neq\mu}}\frac{Z'_\mu}{Z_\mu{-}Z_\alpha}U_{\mu}+(1{-}\delta_{\mu\nu})\frac{Z'_\nu}{Z_\nu{-}Z_\mu}U_{\mu}\;,
\\
\label{eq:Ader}
 	A_{\mu,\nu}^{(i)}=\frac{Z'_\nu}{Z_\nu-Z_\mu}\big(A_{\mu}^{(i)}-A_{\nu}^{(i)}\big)\;,
\\
\label{eq:UdivUmu}
 	\bigg(\frac{U}{U_\mu}\bigg)_{\!\!\!,\mu}=0\;,
\\
\label{eq:DAiAn}
 	 \bigg(\frac{A^{(i)}}{A^{(N)}}\bigg)_{\!\!\!,\mu}=-\frac{Z'_\mu}{Z_\mu}\frac{A_{\mu}^{(i)}}{A^{(N)}}\;.
\end{gather}

%\begin{equation}\label{eq:AUid}
%\begin{split}
%	\sum_\mu \frac{A_{\mu}^{(i)}}{U_\mu} \big(\!-\!Z_\mu\big)^{\!N-1-j} =\delta_{ij}\;, \\
%	\sum_i  \frac{A_{\mu}^{(i)}}{U_\nu}\big(\!-\!Z_\nu\big)^{\!N-1-i} =\delta_{\mu\nu}\;,
%\end{split}
%\end{equation}
%\begin{equation}\label{eq:Uid}
% 	\frac{1}{U_\nu}\sum_{\substack{\mu\\ \mu\neq\nu}}\frac{1}{Z_\mu-Z_\nu}=\sum_{\substack{\mu\\ \mu\neq\nu}}\frac{1}{U_\mu}\frac{1}{Z_\nu-Z_\mu}\;,
%\end{equation}
%\begin{equation}\label{eq:zAAA}
%	A^{(i)}=A_{\mu}^{(i)}+Z_\mu A_{\mu}^{(i-1)}\;,	
%\end{equation}
%\begin{equation}\label{eq:zAUid}
%	\sum_\mu\frac{A_{\mu}^{(j)}}{Z_\mu U_\mu}=\frac{A^{(j)}}{A^{(N)}}\;, 	
%\end{equation}
%\begin{equation}\label{eq:Uder}
% 	U_{\mu,\nu} =\delta_{\mu\nu}\sum_{\substack{\alpha\\ \alpha\neq\mu}}\frac{Z'_\mu}{Z_\mu{-}Z_\alpha}U_{\mu}+(1{-}\delta_{\mu\nu})\frac{Z'_\nu}{Z_\nu{-}Z_\mu}U_{\mu}\;,
%\end{equation}
%\begin{equation}\label{eq:Ader}
% 	A_{\mu,\nu}^{(i)}=\frac{Z'_\nu}{Z_\nu-Z_\mu}\big(A_{\mu}^{(i)}-A_{\nu}^{(i)}\big)\;,
%\end{equation}
%\begin{equation}\label{eq:UdivUmu}
% 	\bigg(\frac{U}{U_\mu}\bigg)_{\!\!\!,\mu}=0\;,
%\end{equation}
%\begin{equation}\label{eq:DAiAn}
% 	 \bigg(\frac{A^{(i)}}{A^{(N)}}\bigg)_{\!\!\!,\mu}=-\frac{Z'_\mu}{Z_\mu}\frac{A_{\mu}^{(i)}}{A^{(N)}}\;.
%\end{equation}
%
%

We state here an important lemma, which in one direction is a consequence of identities \eqref{eq:AUid} and in the other direction involves more delicate arguments, cf.~\cite{Krtous:2007}.
\begin{lemma}\label{le:Fmueq}
Functions ${F_\mu=F_\mu(x_\mu)}$ satisfy equation
\begin{equation}\label{eq:Fmuequation}
	\sum_\mu \frac{F_\mu}{U_\mu} =0\;,
\end{equation}
if and only if they are given by
\begin{equation}\label{eq:Fmuequationsol}
	F_\mu = \sum_{k=0}^{N-2} a_k \big(\!-\!Z_\mu\big)^{\!k}\;,
\end{equation}
where $a_k$ are arbitrary constants.
\end{lemma}

If the right-hand side of equation \eqref{eq:Fmuequation} is nontrivial, the solution of is given by a homogeneous solution \eqref{eq:Fmuequationsol} plus a nonhomogeneous solution. Two special nonhomogeneous solutions, which follow from identities \eqref{eq:AUid} and \eqref{eq:zAUid}, are
\begin{equation}\label{eq:zUid}
	\sum_\mu\frac{\big(\!-\!Z_\mu\big)^{\!N-1}}{U_\mu}=1\;,
	\quad
	\sum_\mu\frac{1}{Z_\mu U_\mu}=\frac{1}{A^{(N)}}\;. 	
\end{equation}

\section{Curvature of the Klein--Gordon separable metric}\label{ap:curv}

From the first structure equations,
\begin{equation}\label{eq:streqI}
	\tb{\dd}\ef{a} +\tb{\omega}[^a_b]\wedge\ef{b}=0\;,
	\quad
	\tb{\omega}[_{ab}] + \tb{\omega}[_{ba}]=0\;,
\end{equation}
%$\tb{\dd}\ef{a} +\tb{\omega}[^a_b]\wedge\ef{b}=0$,
%$\tb{\omega}[_{ab}] + \tb{\omega}[_{ba}]=0$,
and by employing \eqref{eq:Uder} and \eqref{eq:Ader}, we obtain the connection 1-forms
\begin{equation}\label{eq:connectionforms}
\begin{split}
	\tb{\omega}[_{\mu\nu}] &= (1-\delta_{\mu\nu})\frac{1}{2}\frac{1}{Z_\nu-Z_\mu}\bigg(Z'_\nu\sqrt{\frac{X_\nu}{U_{\nu\vphantom{\mu}}}}\ef{\mu}+Z'_\mu\sqrt{\frac{X_\mu}{U_\mu}}\ef{\nu}\bigg)\;,
	\\
	\tb{\omega}[_{\hat{\mu}\hat{\nu}}] &= (1-\delta_{\mu\nu})\frac{1}{2}\frac{1}{Z_\nu-Z_\mu}\bigg(Z'_\mu\sqrt{\frac{X_\nu}{U_{\nu\vphantom{\mu}}}}\ef{\mu}+Z'_\nu\sqrt{\frac{X_\mu}{U_\mu}}\ef{\nu}\bigg)\;,
	\\
	\tb{\omega}[_{\mu\hat{\nu}}] &= \delta_{\mu\nu}\bigg[\!-\bigg(\sqrt{\frac{X_\mu}{U_\mu}}\bigg)_{\!\!,\mu}\efh{\mu}+\frac{1}{2}\sum_{\substack{\rho\\ \rho\neq\mu}}\frac{Z'_\mu}{Z_\mu-Z_\rho}\sqrt{\frac{X_\rho}{U_\rho}}\efh{\rho}\bigg]
	\\
	 &\feq+(1-\delta_{\mu\nu})\frac{1}{2}\frac{Z'_\mu}{Z_\mu-Z_\nu}\bigg(\sqrt{\frac{X_\nu}{U_{\nu\vphantom{\mu}}}}\efh{\mu}-\sqrt{\frac{X_\mu}{U_\mu}}\efh{\nu}\bigg)\;.	 
\end{split} %\raisetag{2ex}
\end{equation}
Straightforward, but very long, calculations of curvature 2-forms $\tb{\Omega}[^a_b]$ from the second structure equations,
%$\tb{\Omega}[^a_b]=\tb{\dd}\tb{\omega}[^a_b]+\tb{\omega}[^a_c]\wedge\tb{\omega}[^c_b]$,
\begin{equation}\label{eq:streqII}
	\tb{\Omega}[^a_b]=\tb{\dd}\tb{\omega}[^a_b]+\tb{\omega}[^a_c]\wedge\tb{\omega}[^c_b]
\end{equation}
lead to the Riemann tensor
%\footnote{%
%Here, for two 2-forms we define $\alpha\varovee\beta=\alpha\beta+\beta\alpha$.}
${\tb{R}=\tb{\Omega}[_{ab}]\ef{a}\ef{b}}$,
\begin{widetext}
\begin{equation}\label{eq:zRiem}
\begin{split}
	\tb{R}&=\sum_{\substack{\mu,\nu\\ \nu\neq\mu}}\frac14 V_{\mu\nu}\eR{\mu}{\nu}{\mu}{\nu}
	+\sum_{\substack{\mu,\nu\\ \nu\neq\mu}}\frac12 W_{\mu\nu}\eRr{\mu}{\nu}{\mu}{\nu}
	\\[-1ex]
	&\feq+\sum_{\substack{\mu,\nu\\ \nu\neq\mu}}\frac14\bigg(V_{\mu\nu}+\frac14\frac{S_{\nu,\mu}}{Z'_\mu}\frac{X_\mu}{U_\mu}+\frac14\frac{S_{\mu,\nu}}{Z'_\nu}\frac{X_\nu}{U_\nu}\bigg)\eRh{\mu}{\nu}{\mu}{\nu}
	+\sum_{\mathclap{\substack{\mu,\nu,\kappa\\ \nu\neq\mu \\ \kappa\neq\mu,\nu}}}\frac12 P_{\mu\nu\kappa}\sqrt{\frac{X_\nu}{U_{\nu\vphantom{\mu}}} \frac{X_\kappa}{U_\kappa}}\eRh{\mu}{\nu}{\mu}{\kappa}
	\\[-1ex]
	&\feq+\sum_{\substack{\mu,\nu\\ \nu\neq\mu}}\frac14 \frac{S_{\nu,\mu}}{Z'_\mu}\sqrt{\frac{X_\mu}{U_\mu} \frac{X_\nu}{U_\nu}}\eRm{\mu}{\mu}{\mu}{\nu}
	+\sum_{\substack{\mu,\nu\\ \nu\neq\mu}}\frac12\bigg(V_{\mu\nu}+\frac14\frac{S_{\mu,\nu}}{Z'_\nu}\frac{X_\nu}{U_\nu}\bigg)\eRm{\mu}{\nu}{\mu}{\nu}
	\\[-1ex]
	&\feq+\sum_{\substack{\mu,\nu\\ \nu\neq\mu}}\frac12 W_{\mu\nu}\eRm{\mu}{\nu}{\nu}{\mu}
	+\sum_{\mathclap{\substack{\mu,\nu,\kappa\\ \nu\neq\mu \\ \kappa\neq\mu,\nu}}}\frac12 P_{\mu\nu\kappa}\sqrt{\frac{X_\nu}{U_{\nu\vphantom{\mu}}}\frac{X_\kappa}{U_\kappa} }\eRm{\mu}{\nu}{\mu}{\kappa}
	\\[-1ex]
	&\feq+\sum_{\substack{\mu,\nu\\ \nu\neq\mu}} W_{\mu\nu}\eRm{\mu}{\mu}{\nu}{\nu}
	 +\sum_\mu\frac14\bigg[-\bigg(\sum_\nu\frac{X_\nu}{U_\nu}\bigg)_{\!\!,\mu\mu}-\sum_{\substack{\nu\\ \nu\neq\mu}}\frac12\frac{S_{\nu,\mu}}{Z'_\mu}\frac{X_\nu}{U_\nu}\bigg]\eRm{\mu}{\mu}{\mu}{\mu}\;.
\end{split}\raisetag{27ex}
\end{equation}
Here, for two 2-forms we, define $\tb{\alpha}\varovee\tb{\beta}=\tb{\alpha}\tb{\beta}+\tb{\beta}\tb{\alpha}$, and
\begin{equation}
\begin{split}
	V_{\mu\nu} &=\frac14\frac{1}{Z_\nu-Z_\mu}\bigg[\frac{Z_\mu}{Z'_\mu}\bigg(\sum_\kappa \frac{{{}Z'_{\kappa}}^2}{Z_\kappa}\frac{X_\kappa}{U_\kappa}\bigg)_{\!\!,\mu}-\frac{Z_\nu}{Z'_\nu}\bigg(\sum_\kappa \frac{{{}Z'_{\kappa}}^2}{Z_\kappa}\frac{X_\kappa}{U_\kappa}\bigg)_{\!\!,\nu}\bigg]\;,
	\\
	W_{\mu\nu} &=\frac14\frac{1}{Z_\nu-Z_\mu}\bigg[Z'_\nu\bigg(\sum_\kappa \frac{X_\kappa}{U_\kappa}\bigg)_{\!\!,\mu}-Z'_\mu\bigg(\sum_\kappa \frac{X_\kappa}{U_\kappa}\bigg)_{\!\!,\nu}\bigg]\;,
	\\[-0.5ex]
	P_{\mu\nu\kappa} &=\frac14\bigg[\frac{{Z'_{\mu}}^2}{(Z_\mu-Z_\nu)(Z_\mu-Z_\kappa)}+\frac{{Z'_{\nu}}^2}{(Z_\nu-Z_\mu)(Z_\nu-Z_\kappa)}+\frac{{Z'_\kappa}^2}{(Z_\kappa-Z_\mu)(Z_\kappa-Z_\nu)}\bigg]\;.
\end{split}
\end{equation}
%\pagebreak[4]
\end{widetext}

%\vfill
%\vspace{20cm}

%%%%%%%%%%%%%%%%%%%%%%%%%%%%%%%%%%%%%%%%%%%%%%%%%%%%%%%%%%%%%%%%%%%%%%%%%%%%%%%%%%%%%%%%%%%%%%%%%%%%%%%%%%%%
%% REFERENCES

%%\bibliographystyle{apsrev4-1} %leave this commented out when using longbibliography option
%\bibliography{references_v203.bib}

\begin{thebibliography}{48}%
\makeatletter
\providecommand \@ifxundefined [1]{%
 \@ifx{#1\undefined}
}%
\providecommand \@ifnum [1]{%
 \ifnum #1\expandafter \@firstoftwo
 \else \expandafter \@secondoftwo
 \fi
}%
\providecommand \@ifx [1]{%
 \ifx #1\expandafter \@firstoftwo
 \else \expandafter \@secondoftwo
 \fi
}%
\providecommand \natexlab [1]{#1}%
\providecommand \enquote  [1]{``#1''}%
\providecommand \bibnamefont  [1]{#1}%
\providecommand \bibfnamefont [1]{#1}%
\providecommand \citenamefont [1]{#1}%
\providecommand \href@noop [0]{\@secondoftwo}%
\providecommand \href [0]{\begingroup \@sanitize@url \@href}%
\providecommand \@href[1]{\@@startlink{#1}\@@href}%
\providecommand \@@href[1]{\endgroup#1\@@endlink}%
\providecommand \@sanitize@url [0]{\catcode `\\12\catcode `\$12\catcode
  `\&12\catcode `\#12\catcode `\^12\catcode `\_12\catcode `\%12\relax}%
\providecommand \@@startlink[1]{}%
\providecommand \@@endlink[0]{}%
\providecommand \url  [0]{\begingroup\@sanitize@url \@url }%
\providecommand \@url [1]{\endgroup\@href {#1}{\urlprefix }}%
\providecommand \urlprefix  [0]{URL }%
\providecommand \Eprint [0]{\href }%
\providecommand \doibase [0]{http://dx.doi.org/}%
\providecommand \selectlanguage [0]{\@gobble}%
\providecommand \bibinfo  [0]{\@secondoftwo}%
\providecommand \bibfield  [0]{\@secondoftwo}%
\providecommand \translation [1]{[#1]}%
\providecommand \BibitemOpen [0]{}%
\providecommand \bibitemStop [0]{}%
\providecommand \bibitemNoStop [0]{.\EOS\space}%
\providecommand \EOS [0]{\spacefactor3000\relax}%
\providecommand \BibitemShut  [1]{\csname bibitem#1\endcsname}%
\let\auto@bib@innerbib\@empty
%</preamble>
\bibitem [{\citenamefont {Kerr}(1963)}]{Kerr:1963}%
  \BibitemOpen
  \bibfield  {author} {\bibinfo {author} {\bibfnamefont {Roy~P.}\ \bibnamefont
  {Kerr}},\ }\bibfield  {title} {\enquote {\bibinfo {title} {Gravitational
  field of a spinning mass as an example of algebraically special metrics},}\
  }\href {\doibase 10.1103/PhysRevLett.11.237} {\bibfield  {journal} {\bibinfo
  {journal} {Phys. Rev. Lett.}\ }\textbf {\bibinfo {volume} {11}},\ \bibinfo
  {pages} {237--238} (\bibinfo {year} {1963})}\BibitemShut {NoStop}%
\bibitem [{\citenamefont {Carter}(1968)}]{Carter:1968b}%
  \BibitemOpen
  \bibfield  {author} {\bibinfo {author} {\bibfnamefont {Brandon}\ \bibnamefont
  {Carter}},\ }\bibfield  {title} {\enquote {\bibinfo {title}
  {{Hamilton-Jacobi} and {Schr\"odinger} separable solutions of {Einstein's}
  equations},}\ }\href@noop {} {\bibfield  {journal} {\bibinfo  {journal}
  {Commun. Math. Phys.}\ }\textbf {\bibinfo {volume} {10}},\ \bibinfo {pages}
  {280--310} (\bibinfo {year} {1968})}\BibitemShut {NoStop}%
\bibitem [{\citenamefont {Carter}(2009)}]{Carter:2009}%
  \BibitemOpen
  \bibfield  {author} {\bibinfo {author} {\bibfnamefont {Brandon}\ \bibnamefont
  {Carter}},\ }\bibfield  {title} {\enquote {\bibinfo {title} {Republication
  of: Black hole equilibrium states},}\ }\href {\doibase
  10.1007/s10714-009-0888-5} {\bibfield  {journal} {\bibinfo  {journal}
  {General Relativity and Gravitation}\ }\textbf {\bibinfo {volume} {41}},\
  \bibinfo {pages} {2873--2938} (\bibinfo {year} {2009})}\BibitemShut {NoStop}%
\bibitem [{\citenamefont {Tangherlini}(1963)}]{Tangherlini:1963}%
  \BibitemOpen
  \bibfield  {author} {\bibinfo {author} {\bibfnamefont {F.~R.}\ \bibnamefont
  {Tangherlini}},\ }\bibfield  {title} {\enquote {\bibinfo {title}
  {Schwartzschild field in $n$ dimensions and the dimensionality of space
  problem},}\ }\href@noop {} {\bibfield  {journal} {\bibinfo  {journal} {Nuovo
  Cimento}\ }\textbf {\bibinfo {volume} {27}},\ \bibinfo {pages} {636}
  (\bibinfo {year} {1963})}\BibitemShut {NoStop}%
\bibitem [{\citenamefont {Myers}\ and\ \citenamefont
  {Perry}(1986)}]{MyersPerry:1986}%
  \BibitemOpen
  \bibfield  {author} {\bibinfo {author} {\bibfnamefont {R.~C.}\ \bibnamefont
  {Myers}}\ and\ \bibinfo {author} {\bibfnamefont {M.~J.}\ \bibnamefont
  {Perry}},\ }\bibfield  {title} {\enquote {\bibinfo {title} {Black holes in
  higher dimensional space-times},}\ }\href@noop {} {\bibfield  {journal}
  {\bibinfo  {journal} {Ann. Phys. (N.Y.)}\ }\textbf {\bibinfo {volume}
  {172}},\ \bibinfo {pages} {304--347} (\bibinfo {year} {1986})}\BibitemShut
  {NoStop}%
\bibitem [{\citenamefont {Hawking}\ \emph {et~al.}(1999)\citenamefont
  {Hawking}, \citenamefont {Hunter},\ and\ \citenamefont
  {Taylor-Robinson}}]{HawkingEtal:1999}%
  \BibitemOpen
  \bibfield  {author} {\bibinfo {author} {\bibfnamefont {S.~W.}\ \bibnamefont
  {Hawking}}, \bibinfo {author} {\bibfnamefont {C.~J.}\ \bibnamefont {Hunter}},
  \ and\ \bibinfo {author} {\bibfnamefont {M.~M.}\ \bibnamefont
  {Taylor-Robinson}},\ }\bibfield  {title} {\enquote {\bibinfo {title}
  {Rotation and the {AdS/CFT} correspondence},}\ }\href@noop {} {\bibfield
  {journal} {\bibinfo  {journal} {Phys. Rev. D}\ }\textbf {\bibinfo {volume}
  {59}},\ \bibinfo {pages} {064005} (\bibinfo {year} {1999})},\ \Eprint
  {http://arxiv.org/abs/arXiv: hep-th/9811056} {arXiv: hep-th/9811056}
  \BibitemShut {NoStop}%
\bibitem [{\citenamefont {Gibbons}\ \emph {et~al.}(2004)\citenamefont
  {Gibbons}, \citenamefont {L\"u}, \citenamefont {Page},\ and\ \citenamefont
  {Pope}}]{GibbonsEtal:2004}%
  \BibitemOpen
  \bibfield  {author} {\bibinfo {author} {\bibfnamefont {G.~W.}\ \bibnamefont
  {Gibbons}}, \bibinfo {author} {\bibfnamefont {H.}~\bibnamefont {L\"u}},
  \bibinfo {author} {\bibfnamefont {Don~N.}\ \bibnamefont {Page}}, \ and\
  \bibinfo {author} {\bibfnamefont {C.~N.}\ \bibnamefont {Pope}},\ }\bibfield
  {title} {\enquote {\bibinfo {title} {Rotating black holes in higher
  dimensions with a cosmological constant},}\ }\href@noop {} {\bibfield
  {journal} {\bibinfo  {journal} {Phys. Rev. Lett.}\ }\textbf {\bibinfo
  {volume} {93}},\ \bibinfo {pages} {171102} (\bibinfo {year} {2004})},\
  \Eprint {http://arxiv.org/abs/arXiv: hep-th/0409155} {arXiv: hep-th/0409155}
  \BibitemShut {NoStop}%
\bibitem [{\citenamefont {Gibbons}\ \emph {et~al.}(2005)\citenamefont
  {Gibbons}, \citenamefont {L\"u}, \citenamefont {Page},\ and\ \citenamefont
  {Pope}}]{GibbonsEtal:2005}%
  \BibitemOpen
  \bibfield  {author} {\bibinfo {author} {\bibfnamefont {G.~W.}\ \bibnamefont
  {Gibbons}}, \bibinfo {author} {\bibfnamefont {H.}~\bibnamefont {L\"u}},
  \bibinfo {author} {\bibfnamefont {Don~N.}\ \bibnamefont {Page}}, \ and\
  \bibinfo {author} {\bibfnamefont {C.~N.}\ \bibnamefont {Pope}},\ }\bibfield
  {title} {\enquote {\bibinfo {title} {The general {K}err-de {S}itter metrics
  in all dimensions},}\ }\href@noop {} {\bibfield  {journal} {\bibinfo
  {journal} {J. Geom. Phys.}\ }\textbf {\bibinfo {volume} {53}},\ \bibinfo
  {pages} {49--73} (\bibinfo {year} {2005})},\ \Eprint
  {http://arxiv.org/abs/arXiv: hep-th/0404008} {arXiv: hep-th/0404008}
  \BibitemShut {NoStop}%
\bibitem [{\citenamefont {Chen}\ \emph
  {et~al.}(2006{\natexlab{a}})\citenamefont {Chen}, \citenamefont {L\"u},\ and\
  \citenamefont {Pope}}]{ChenLuPope:2006}%
  \BibitemOpen
  \bibfield  {author} {\bibinfo {author} {\bibfnamefont {W.}~\bibnamefont
  {Chen}}, \bibinfo {author} {\bibfnamefont {H.}~\bibnamefont {L\"u}}, \ and\
  \bibinfo {author} {\bibfnamefont {C.~N.}\ \bibnamefont {Pope}},\ }\bibfield
  {title} {\enquote {\bibinfo {title} {General {Kerr-NUT-AdS} metrics in all
  dimensions},}\ }\href@noop {} {\bibfield  {journal} {\bibinfo  {journal}
  {Class. Quantum Grav.}\ }\textbf {\bibinfo {volume} {23}},\ \bibinfo {pages}
  {5323--5340} (\bibinfo {year} {2006}{\natexlab{a}})},\ \Eprint
  {http://arxiv.org/abs/arXiv: hep-th/0604125} {arXiv: hep-th/0604125}
  \BibitemShut {NoStop}%
\bibitem [{\citenamefont {Chen}\ \emph {et~al.}(2007)\citenamefont {Chen},
  \citenamefont {L\"u},\ and\ \citenamefont {Pope}}]{ChenLuPope:2007}%
  \BibitemOpen
  \bibfield  {author} {\bibinfo {author} {\bibfnamefont {W.}~\bibnamefont
  {Chen}}, \bibinfo {author} {\bibfnamefont {H.}~\bibnamefont {L\"u}}, \ and\
  \bibinfo {author} {\bibfnamefont {C.~N.}\ \bibnamefont {Pope}},\ }\bibfield
  {title} {\enquote {\bibinfo {title} {{Kerr-de Sitter} black holes with {NUT}
  charges},}\ }\href@noop {} {\bibfield  {journal} {\bibinfo  {journal} {Nucl.
  Phys.}\ }\textbf {\bibinfo {volume} {B762}},\ \bibinfo {pages} {38--54}
  (\bibinfo {year} {2007})},\ \Eprint {http://arxiv.org/abs/arXiv:
  hep-th/0601002} {arXiv: hep-th/0601002} \BibitemShut {NoStop}%
\bibitem [{\citenamefont {Page}\ \emph {et~al.}(2007)\citenamefont {Page},
  \citenamefont {Kubiz\v{n}\'ak}, \citenamefont {Vasudevan},\ and\
  \citenamefont {Krtou\v{s}}}]{PageEtal:2007}%
  \BibitemOpen
  \bibfield  {author} {\bibinfo {author} {\bibfnamefont {Don~N.}\ \bibnamefont
  {Page}}, \bibinfo {author} {\bibfnamefont {David}\ \bibnamefont
  {Kubiz\v{n}\'ak}}, \bibinfo {author} {\bibfnamefont {Muraari}\ \bibnamefont
  {Vasudevan}}, \ and\ \bibinfo {author} {\bibfnamefont {Pavel}\ \bibnamefont
  {Krtou\v{s}}},\ }\bibfield  {title} {\enquote {\bibinfo {title} {Complete
  integrability of geodesic motion in general higher-dimensional rotating black
  hole spacetimes},}\ }\href@noop {} {\bibfield  {journal} {\bibinfo  {journal}
  {Phys. Rev. Lett.}\ }\textbf {\bibinfo {volume} {98}},\ \bibinfo {pages}
  {061102} (\bibinfo {year} {2007})},\ \Eprint {http://arxiv.org/abs/arXiv:
  hep-th/0611083} {arXiv: hep-th/0611083} \BibitemShut {NoStop}%
\bibitem [{\citenamefont {Frolov}\ \emph {et~al.}(2007)\citenamefont {Frolov},
  \citenamefont {Krtou\v{s}},\ and\ \citenamefont
  {Kubiz\v{n}\'ak}}]{FrolovEtal:2007}%
  \BibitemOpen
  \bibfield  {author} {\bibinfo {author} {\bibfnamefont {Valeri~P.}\
  \bibnamefont {Frolov}}, \bibinfo {author} {\bibfnamefont {Pavel}\
  \bibnamefont {Krtou\v{s}}}, \ and\ \bibinfo {author} {\bibfnamefont {David}\
  \bibnamefont {Kubiz\v{n}\'ak}},\ }\bibfield  {title} {\enquote {\bibinfo
  {title} {Separability of {Hamilton--Jacobi} and {Klein--Gordon} equations in
  general {Kerr-NUT-AdS} spacetimes},}\ }\href@noop {} {\bibfield  {journal}
  {\bibinfo  {journal} {J. High Energy Phys.}\ }\textbf {\bibinfo {volume}
  {0702}},\ \bibinfo {pages} {005} (\bibinfo {year} {2007})},\ \Eprint
  {http://arxiv.org/abs/arXiv: hep-th/0611245} {arXiv: hep-th/0611245}
  \BibitemShut {NoStop}%
\bibitem [{\citenamefont {Sergyeyev}\ and\ \citenamefont
  {Krtou\v{s}}(2008)}]{SergyeyevKrtous:2008}%
  \BibitemOpen
  \bibfield  {author} {\bibinfo {author} {\bibfnamefont {Artur}\ \bibnamefont
  {Sergyeyev}}\ and\ \bibinfo {author} {\bibfnamefont {Pavel}\ \bibnamefont
  {Krtou\v{s}}},\ }\bibfield  {title} {\enquote {\bibinfo {title} {Complete set
  of commuting symmetry operators for the {Klein-Gordon} equation in
  generalized higher-dimensional {Kerr-NUT-(A)dS} spacetimes},}\ }\href@noop {}
  {\bibfield  {journal} {\bibinfo  {journal} {Phys. Rev. D}\ }\textbf {\bibinfo
  {volume} {77}},\ \bibinfo {pages} {044033} (\bibinfo {year} {2008})},\
  \Eprint {http://arxiv.org/abs/arXiv: 0711.4623 [hep-th]} {arXiv: 0711.4623
  [hep-th]} \BibitemShut {NoStop}%
\bibitem [{\citenamefont {Cariglia}\ \emph
  {et~al.}(2011{\natexlab{a}})\citenamefont {Cariglia}, \citenamefont
  {Krtou\v{s}},\ and\ \citenamefont {Kubiz\v{n}\'ak}}]{CarigliaEtal:2011a}%
  \BibitemOpen
  \bibfield  {author} {\bibinfo {author} {\bibfnamefont {Marco}\ \bibnamefont
  {Cariglia}}, \bibinfo {author} {\bibfnamefont {Pavel}\ \bibnamefont
  {Krtou\v{s}}}, \ and\ \bibinfo {author} {\bibfnamefont {David}\ \bibnamefont
  {Kubiz\v{n}\'ak}},\ }\bibfield  {title} {\enquote {\bibinfo {title}
  {Commuting symmetry operators of the {Dirac} equation, {Killing-Yano} and
  {Schouten-Nijenhuis} brackets},}\ }\href {\doibase
  10.1103/PhysRevD.84.024004} {\bibfield  {journal} {\bibinfo  {journal} {Phys.
  Rev. D}\ }\textbf {\bibinfo {volume} {84}},\ \bibinfo {pages} {024004}
  (\bibinfo {year} {2011}{\natexlab{a}})},\ \Eprint
  {http://arxiv.org/abs/arXiv: 1102.4501 [hep-th]} {arXiv: 1102.4501 [hep-th]}
  \BibitemShut {NoStop}%
\bibitem [{\citenamefont {Cariglia}\ \emph
  {et~al.}(2011{\natexlab{b}})\citenamefont {Cariglia}, \citenamefont
  {Krtou\v{s}},\ and\ \citenamefont {Kubiz\v{n}\'ak}}]{CarigliaEtal:2011b}%
  \BibitemOpen
  \bibfield  {author} {\bibinfo {author} {\bibfnamefont {Marco}\ \bibnamefont
  {Cariglia}}, \bibinfo {author} {\bibfnamefont {Pavel}\ \bibnamefont
  {Krtou\v{s}}}, \ and\ \bibinfo {author} {\bibfnamefont {David}\ \bibnamefont
  {Kubiz\v{n}\'ak}},\ }\bibfield  {title} {\enquote {\bibinfo {title} {{Dirac}
  equation in {Kerr-NUT-(A)dS} spacetimes: Intrinsic characterization of
  separability in all dimensions},}\ }\href {\doibase
  10.1103/PhysRevD.84.024008} {\bibfield  {journal} {\bibinfo  {journal} {Phys.
  Rev. D}\ }\textbf {\bibinfo {volume} {84}},\ \bibinfo {pages} {024008}
  (\bibinfo {year} {2011}{\natexlab{b}})},\ \Eprint
  {http://arxiv.org/abs/arXiv: 1104.4123 [hep-th]} {arXiv: 1104.4123 [hep-th]}
  \BibitemShut {NoStop}%
\bibitem [{\citenamefont {Krtou\v{s}}\ \emph
  {et~al.}(2007{\natexlab{a}})\citenamefont {Krtou\v{s}}, \citenamefont
  {Kubiz\v{n}\'ak}, \citenamefont {Page},\ and\ \citenamefont
  {Frolov}}]{KrtousEtal:2007a}%
  \BibitemOpen
  \bibfield  {author} {\bibinfo {author} {\bibfnamefont {Pavel}\ \bibnamefont
  {Krtou\v{s}}}, \bibinfo {author} {\bibfnamefont {David}\ \bibnamefont
  {Kubiz\v{n}\'ak}}, \bibinfo {author} {\bibfnamefont {Don~N.}\ \bibnamefont
  {Page}}, \ and\ \bibinfo {author} {\bibfnamefont {Valeri~P.}\ \bibnamefont
  {Frolov}},\ }\bibfield  {title} {\enquote {\bibinfo {title} {{Killing--Yano}
  tensors, rank-2 {Killing} tensors, and conserved quantities in higher
  dimensions},}\ }\href@noop {} {\bibfield  {journal} {\bibinfo  {journal} {J.
  High Energy Phys.}\ }\textbf {\bibinfo {volume} {0702}},\ \bibinfo {pages}
  {004} (\bibinfo {year} {2007}{\natexlab{a}})},\ \Eprint
  {http://arxiv.org/abs/arXiv: hep-th/0612029} {arXiv: hep-th/0612029}
  \BibitemShut {NoStop}%
\bibitem [{\citenamefont {Krtou\v{s}}\ \emph
  {et~al.}(2007{\natexlab{b}})\citenamefont {Krtou\v{s}}, \citenamefont
  {Kubiz\v{n}\'ak}, \citenamefont {Page},\ and\ \citenamefont
  {Vasudevan}}]{KrtousEtal:2007b}%
  \BibitemOpen
  \bibfield  {author} {\bibinfo {author} {\bibfnamefont {Pavel}\ \bibnamefont
  {Krtou\v{s}}}, \bibinfo {author} {\bibfnamefont {David}\ \bibnamefont
  {Kubiz\v{n}\'ak}}, \bibinfo {author} {\bibfnamefont {Don~N.}\ \bibnamefont
  {Page}}, \ and\ \bibinfo {author} {\bibfnamefont {Muraari}\ \bibnamefont
  {Vasudevan}},\ }\bibfield  {title} {\enquote {\bibinfo {title} {Constants of
  geodesic motion in higher-dimensional black-hole spacetimes},}\ }\href@noop
  {} {\bibfield  {journal} {\bibinfo  {journal} {Phys. Rev. D}\ }\textbf
  {\bibinfo {volume} {76}},\ \bibinfo {pages} {084034} (\bibinfo {year}
  {2007}{\natexlab{b}})},\ \Eprint {http://arxiv.org/abs/arXiv: 0707.0001
  [hep-th]} {arXiv: 0707.0001 [hep-th]} \BibitemShut {NoStop}%
\bibitem [{\citenamefont {Walker}\ and\ \citenamefont
  {Penrose}(1970)}]{WalkerPenrose:1970}%
  \BibitemOpen
  \bibfield  {author} {\bibinfo {author} {\bibfnamefont {Martin}\ \bibnamefont
  {Walker}}\ and\ \bibinfo {author} {\bibfnamefont {Roger}\ \bibnamefont
  {Penrose}},\ }\bibfield  {title} {\enquote {\bibinfo {title} {On quadratic
  first integrals of the geodesic equations for type \{22\} spacetimes},}\
  }\href@noop {} {\bibfield  {journal} {\bibinfo  {journal} {Commun. Math.
  Phys.}\ }\textbf {\bibinfo {volume} {18}},\ \bibinfo {pages} {265--274}
  (\bibinfo {year} {1970})}\BibitemShut {NoStop}%
\bibitem [{\citenamefont {Teukolsky}(1972)}]{Teukolsky:1972}%
  \BibitemOpen
  \bibfield  {author} {\bibinfo {author} {\bibfnamefont {Saul~A.}\ \bibnamefont
  {Teukolsky}},\ }\bibfield  {title} {\enquote {\bibinfo {title} {{Rotating
  black holes - separable wave equations for gravitational and electromagnetic
  perturbations}},}\ }\href {\doibase 10.1103/PhysRevLett.29.1114} {\bibfield
  {journal} {\bibinfo  {journal} {Phys. Rev. Lett.}\ }\textbf {\bibinfo
  {volume} {29}},\ \bibinfo {pages} {1114--1118} (\bibinfo {year}
  {1972})}\BibitemShut {NoStop}%
\bibitem [{\citenamefont {Teukolsky}(1973)}]{Teukolsky:1973}%
  \BibitemOpen
  \bibfield  {author} {\bibinfo {author} {\bibfnamefont {Saul~A.}\ \bibnamefont
  {Teukolsky}},\ }\bibfield  {title} {\enquote {\bibinfo {title} {Perturbations
  of a rotating black hole. 1. fundamental equations for gravitational
  electromagnetic and neutrino field perturbations},}\ }\href {\doibase
  10.1086/152444} {\bibfield  {journal} {\bibinfo  {journal} {Astrophys. J.}\
  }\textbf {\bibinfo {volume} {185}},\ \bibinfo {pages} {635--647} (\bibinfo
  {year} {1973})}\BibitemShut {NoStop}%
\bibitem [{\citenamefont {Carter}(1977)}]{Carter:1977}%
  \BibitemOpen
  \bibfield  {author} {\bibinfo {author} {\bibfnamefont {Brandon}\ \bibnamefont
  {Carter}},\ }\bibfield  {title} {\enquote {\bibinfo {title} {Killing tensor
  quantum numbers and conserved currents in curved space},}\ }\href {\doibase
  10.1103/PhysRevD.16.3395} {\bibfield  {journal} {\bibinfo  {journal} {Phys.
  Rev. D}\ }\textbf {\bibinfo {volume} {16}},\ \bibinfo {pages} {3395--3414}
  (\bibinfo {year} {1977})}\BibitemShut {NoStop}%
\bibitem [{\citenamefont {Demianski}\ and\ \citenamefont
  {Francaviglia}(1981)}]{DemianskiFrancaviglia:1981}%
  \BibitemOpen
  \bibfield  {author} {\bibinfo {author} {\bibfnamefont {M.}~\bibnamefont
  {Demianski}}\ and\ \bibinfo {author} {\bibfnamefont {M.}~\bibnamefont
  {Francaviglia}},\ }\bibfield  {title} {\enquote {\bibinfo {title} {{Type-D}
  space-times with a {Killing} tensor},}\ }\href@noop {} {\bibfield  {journal}
  {\bibinfo  {journal} {Journal of Physics A: Mathematical and General}\
  }\textbf {\bibinfo {volume} {14}},\ \bibinfo {pages} {173} (\bibinfo {year}
  {1981})}\BibitemShut {NoStop}%
\bibitem [{\citenamefont {Frolov}\ and\ \citenamefont
  {Stojkovi\'{c}}(2003{\natexlab{a}})}]{FrolovStojkovic:2003a}%
  \BibitemOpen
  \bibfield  {author} {\bibinfo {author} {\bibfnamefont {Valeri}\ \bibnamefont
  {Frolov}}\ and\ \bibinfo {author} {\bibfnamefont {Dejan}\ \bibnamefont
  {Stojkovi\'{c}}},\ }\bibfield  {title} {\enquote {\bibinfo {title} {Particle
  and light motion in a space-time of a five-dimensional rotating black
  hole},}\ }\href@noop {} {\bibfield  {journal} {\bibinfo  {journal} {Phys.
  Rev. D}\ }\textbf {\bibinfo {volume} {68}},\ \bibinfo {pages} {064011}
  (\bibinfo {year} {2003}{\natexlab{a}})},\ \Eprint
  {http://arxiv.org/abs/arXiv: gr-qc/0301016} {arXiv: gr-qc/0301016}
  \BibitemShut {NoStop}%
\bibitem [{\citenamefont {Frolov}\ and\ \citenamefont
  {Stojkovi\'{c}}(2003{\natexlab{b}})}]{FrolovStojkovic:2003b}%
  \BibitemOpen
  \bibfield  {author} {\bibinfo {author} {\bibfnamefont {Valeri}\ \bibnamefont
  {Frolov}}\ and\ \bibinfo {author} {\bibfnamefont {Dejan}\ \bibnamefont
  {Stojkovi\'{c}}},\ }\bibfield  {title} {\enquote {\bibinfo {title} {Quantum
  radiation from a $5$-dimensional rotating black hole},}\ }\href {\doibase
  10.1103/PhysRevD.67.084004} {\bibfield  {journal} {\bibinfo  {journal} {Phys.
  Rev. D}\ }\textbf {\bibinfo {volume} {67}},\ \bibinfo {pages} {084004}
  (\bibinfo {year} {2003}{\natexlab{b}})},\ \Eprint
  {http://arxiv.org/abs/arXiv: gr-qc/0211055} {arXiv: gr-qc/0211055}
  \BibitemShut {NoStop}%
\bibitem [{\citenamefont {Chong}\ \emph {et~al.}(2005)\citenamefont {Chong},
  \citenamefont {Gibbons}, \citenamefont {Lu},\ and\ \citenamefont
  {Pope}}]{ChongEtal:2004}%
  \BibitemOpen
  \bibfield  {author} {\bibinfo {author} {\bibfnamefont {Z.~W.}\ \bibnamefont
  {Chong}}, \bibinfo {author} {\bibfnamefont {G.~W.}\ \bibnamefont {Gibbons}},
  \bibinfo {author} {\bibfnamefont {H.}~\bibnamefont {Lu}}, \ and\ \bibinfo
  {author} {\bibfnamefont {C.~N.}\ \bibnamefont {Pope}},\ }\bibfield  {title}
  {\enquote {\bibinfo {title} {Separability and {Killing} tensors in
  {Kerr-Taub-NUT-de Sitter} metrics in higher dimensions},}\ }\href {\doibase
  10.1016/j.physletb.2004.07.066} {\bibfield  {journal} {\bibinfo  {journal}
  {Phys. Lett.}\ }\textbf {\bibinfo {volume} {B609}},\ \bibinfo {pages}
  {124--132} (\bibinfo {year} {2005})},\ \Eprint {http://arxiv.org/abs/arXiv:
  hep-th/0405061} {arXiv: hep-th/0405061} \BibitemShut {NoStop}%
\bibitem [{\citenamefont {Vasudevan}\ \emph
  {et~al.}(2005{\natexlab{a}})\citenamefont {Vasudevan}, \citenamefont
  {Stevens},\ and\ \citenamefont {Page}}]{VasudevanEtal:2004}%
  \BibitemOpen
  \bibfield  {author} {\bibinfo {author} {\bibfnamefont {Muraari}\ \bibnamefont
  {Vasudevan}}, \bibinfo {author} {\bibfnamefont {Kory~A.}\ \bibnamefont
  {Stevens}}, \ and\ \bibinfo {author} {\bibfnamefont {Don~N.}\ \bibnamefont
  {Page}},\ }\bibfield  {title} {\enquote {\bibinfo {title} {Separability of
  the {Hamilton-Jacobi} and {Klein-Gordon} equations in {Kerr-de Sitter}
  metrics},}\ }\href {\doibase 10.1088/0264-9381/22/2/007} {\bibfield
  {journal} {\bibinfo  {journal} {Class. Quantum Grav.}\ }\textbf {\bibinfo
  {volume} {22}},\ \bibinfo {pages} {339--352} (\bibinfo {year}
  {2005}{\natexlab{a}})},\ \Eprint {http://arxiv.org/abs/arXiv: gr-qc/0405125}
  {arXiv: gr-qc/0405125} \BibitemShut {NoStop}%
\bibitem [{\citenamefont {Kunduri}\ and\ \citenamefont
  {Lucietti}(2005)}]{KunduriLucietti:2005}%
  \BibitemOpen
  \bibfield  {author} {\bibinfo {author} {\bibfnamefont {Hari~K.}\ \bibnamefont
  {Kunduri}}\ and\ \bibinfo {author} {\bibfnamefont {James}\ \bibnamefont
  {Lucietti}},\ }\bibfield  {title} {\enquote {\bibinfo {title} {Integrability
  and the {Kerr-(A)dS} black hole in five dimensions},}\ }\href {\doibase
  arXiv: 10.1103/PhysRevD.71.104021} {\bibfield  {journal} {\bibinfo  {journal}
  {Phys. Rev. D}\ }\textbf {\bibinfo {volume} {71}},\ \bibinfo {pages} {104021}
  (\bibinfo {year} {2005})},\ \Eprint {http://arxiv.org/abs/hep-th/0502124}
  {hep-th/0502124} \BibitemShut {NoStop}%
\bibitem [{\citenamefont {Vasudevan}\ \emph
  {et~al.}(2005{\natexlab{b}})\citenamefont {Vasudevan}, \citenamefont
  {Stevens},\ and\ \citenamefont {Page}}]{VasudevanEtal:2005}%
  \BibitemOpen
  \bibfield  {author} {\bibinfo {author} {\bibfnamefont {Muraari}\ \bibnamefont
  {Vasudevan}}, \bibinfo {author} {\bibfnamefont {Kory~A.}\ \bibnamefont
  {Stevens}}, \ and\ \bibinfo {author} {\bibfnamefont {Don~N.}\ \bibnamefont
  {Page}},\ }\bibfield  {title} {\enquote {\bibinfo {title} {Particle motion
  and scalar field propagation in {Myers-Perry} black hole spacetimes in all
  dimensions},}\ }\href {\doibase 10.1088/0264-9381/22/7/017} {\bibfield
  {journal} {\bibinfo  {journal} {Class. Quantum Grav.}\ }\textbf {\bibinfo
  {volume} {22}},\ \bibinfo {pages} {1469--1482} (\bibinfo {year}
  {2005}{\natexlab{b}})},\ \Eprint {http://arxiv.org/abs/arXiv: gr-qc/0407030}
  {arXiv: gr-qc/0407030} \BibitemShut {NoStop}%
\bibitem [{\citenamefont {Vasudevan}\ and\ \citenamefont
  {Stevens}(2005)}]{VasudevanStevens:2005}%
  \BibitemOpen
  \bibfield  {author} {\bibinfo {author} {\bibfnamefont {Muraari}\ \bibnamefont
  {Vasudevan}}\ and\ \bibinfo {author} {\bibfnamefont {Kory~A.}\ \bibnamefont
  {Stevens}},\ }\bibfield  {title} {\enquote {\bibinfo {title} {Integrability
  of particle motion and scalar field propagation in {Kerr-(Anti) de Sitter}
  black hole spacetimes in all dimensions},}\ }\href {\doibase
  10.1103/PhysRevD.72.124008} {\bibfield  {journal} {\bibinfo  {journal} {Phys.
  Rev. D}\ }\textbf {\bibinfo {volume} {72}},\ \bibinfo {pages} {124008}
  (\bibinfo {year} {2005})},\ \Eprint {http://arxiv.org/abs/arXiv:
  gr-qc/0507096} {arXiv: gr-qc/0507096} \BibitemShut {NoStop}%
\bibitem [{\citenamefont {Chen}\ \emph
  {et~al.}(2006{\natexlab{b}})\citenamefont {Chen}, \citenamefont {L\"u},\ and\
  \citenamefont {Pope}}]{ChenLuPope:2006b}%
  \BibitemOpen
  \bibfield  {author} {\bibinfo {author} {\bibfnamefont {W.}~\bibnamefont
  {Chen}}, \bibinfo {author} {\bibfnamefont {H.}~\bibnamefont {L\"u}}, \ and\
  \bibinfo {author} {\bibfnamefont {C.~N.}\ \bibnamefont {Pope}},\ }\bibfield
  {title} {\enquote {\bibinfo {title} {Separability in cohomogeneity-2
  {Kerr-NUT-AdS} metrics},}\ }\href {\doibase 10.1088/1126-6708/2006/04/008}
  {\bibfield  {journal} {\bibinfo  {journal} {J. High Energy Phys.}\ }\textbf
  {\bibinfo {volume} {0604}},\ \bibinfo {pages} {008} (\bibinfo {year}
  {2006}{\natexlab{b}})},\ \Eprint {http://arxiv.org/abs/arXiv: hep-th/0602084}
  {arXiv: hep-th/0602084} \BibitemShut {NoStop}%
\bibitem [{\citenamefont {Davis}(2006)}]{Davis:2006}%
  \BibitemOpen
  \bibfield  {author} {\bibinfo {author} {\bibfnamefont {Paul}\ \bibnamefont
  {Davis}},\ }\bibfield  {title} {\enquote {\bibinfo {title} {A {Killing}
  tensor for higher dimensional {Kerr-AdS} black holes with {NUT} charge},}\
  }\href {\doibase 10.1088/0264-9381/23/10/023} {\bibfield  {journal} {\bibinfo
   {journal} {Class. Quantum Grav.}\ }\textbf {\bibinfo {volume} {23}},\
  \bibinfo {pages} {3607--3618} (\bibinfo {year} {2006})},\ \Eprint
  {http://arxiv.org/abs/arXiv: hep-th/0602118} {arXiv: hep-th/0602118}
  \BibitemShut {NoStop}%
\bibitem [{\citenamefont {Frolov}\ and\ \citenamefont
  {Kubiz\v{n}\'ak}(2007)}]{FrolovKubiznak:2007}%
  \BibitemOpen
  \bibfield  {author} {\bibinfo {author} {\bibfnamefont {Valeri~P.}\
  \bibnamefont {Frolov}}\ and\ \bibinfo {author} {\bibfnamefont {David}\
  \bibnamefont {Kubiz\v{n}\'ak}},\ }\bibfield  {title} {\enquote {\bibinfo
  {title} {{`Hidden'} symmetries of higher dimensional rotating black holes},}\
  }\href@noop {} {\bibfield  {journal} {\bibinfo  {journal} {Phys. Rev. Lett.}\
  }\textbf {\bibinfo {volume} {98}},\ \bibinfo {pages} {011101} (\bibinfo {year}
  {2007})},\ \Eprint {http://arxiv.org/abs/arXiv: gr-qc/0605058} {arXiv:
  gr-qc/0605058} \BibitemShut {NoStop}%
\bibitem [{\citenamefont {Kubiz\v{n}\'ak}\ and\ \citenamefont
  {Frolov}(2007)}]{KubiznakFrolov:2007}%
  \BibitemOpen
  \bibfield  {author} {\bibinfo {author} {\bibfnamefont {David}\ \bibnamefont
  {Kubiz\v{n}\'ak}}\ and\ \bibinfo {author} {\bibfnamefont {Valri~P.}\
  \bibnamefont {Frolov}},\ }\bibfield  {title} {\enquote {\bibinfo {title}
  {Hidden symmetry of higher dimensional {Kerr-NUT-AdS} spacetimes},}\
  }\href@noop {} {\bibfield  {journal} {\bibinfo  {journal} {Class. Quantum
  Grav.}\ }\textbf {\bibinfo {volume} {24}},\ \bibinfo {pages} {F1--F6}
  (\bibinfo {year} {2007})},\ \Eprint {http://arxiv.org/abs/arXiv:
  gr-qc/0610144} {arXiv: gr-qc/0610144} \BibitemShut {NoStop}%
\bibitem [{\citenamefont {Benenti}\ and\ \citenamefont
  {Francaviglia}(1979)}]{BenentiFrancaviglia:1979}%
  \BibitemOpen
  \bibfield  {author} {\bibinfo {author} {\bibfnamefont {S.}~\bibnamefont
  {Benenti}}\ and\ \bibinfo {author} {\bibfnamefont {M.}~\bibnamefont
  {Francaviglia}},\ }\bibfield  {title} {\enquote {\bibinfo {title} {Remarks on
  certain separability structures and their applications to general
  relativity},}\ }\href@noop {} {\bibfield  {journal} {\bibinfo  {journal}
  {Gen. Rel. Grav.}\ }\textbf {\bibinfo {volume} {10}},\ \bibinfo {pages}
  {79--92} (\bibinfo {year} {1979})}\BibitemShut {NoStop}%
\bibitem [{\citenamefont {Benenti}\ and\ \citenamefont
  {Francaviglia}(1980)}]{BenentiFrancaviglia:1980}%
  \BibitemOpen
  \bibfield  {author} {\bibinfo {author} {\bibfnamefont {S.}~\bibnamefont
  {Benenti}}\ and\ \bibinfo {author} {\bibfnamefont {M.}~\bibnamefont
  {Francaviglia}},\ }\bibfield  {title} {\enquote {\bibinfo {title} {The theory
  of separability of the {Hamilton-Jacobi} equation and its applications to
  general relativity},}\ }in\ \href@noop {} {\emph {\bibinfo {booktitle}
  {General Relativity and Gravitation, vol. I}}},\ \bibinfo {editor} {edited
  by\ \bibinfo {editor} {\bibfnamefont {A.}~\bibnamefont {Held}}}\ (\bibinfo
  {publisher} {Plenum Press},\ \bibinfo {address} {New York},\ \bibinfo {year}
  {1980})\ \bibinfo {note} {ch. 13}\BibitemShut {NoStop}%
\bibitem [{\citenamefont {Houri}\ \emph {et~al.}(2007)\citenamefont {Houri},
  \citenamefont {Oota},\ and\ \citenamefont {Yasui}}]{HouriEtal:2007}%
  \BibitemOpen
  \bibfield  {author} {\bibinfo {author} {\bibfnamefont {Tsuyoshi}\
  \bibnamefont {Houri}}, \bibinfo {author} {\bibfnamefont {Takeshi}\
  \bibnamefont {Oota}}, \ and\ \bibinfo {author} {\bibfnamefont {Yukinori}\
  \bibnamefont {Yasui}},\ }\bibfield  {title} {\enquote {\bibinfo {title}
  {Closed conformal {Killing--Yano} tensor and {Kerr-NUT-de Sitter} spacetime
  uniqueness},}\ }\href@noop {} {\bibfield  {journal} {\bibinfo  {journal}
  {Phys. Lett.}\ }\textbf {\bibinfo {volume} {B656}},\ \bibinfo {pages}
  {214--216} (\bibinfo {year} {2007})},\ \Eprint {http://arxiv.org/abs/arXiv:
  0708.1368 [hep-th]} {arXiv: 0708.1368 [hep-th]} \BibitemShut {NoStop}%
\bibitem [{\citenamefont {Krtou{\v s}}\ \emph {et~al.}(2008)\citenamefont
  {Krtou{\v s}}, \citenamefont {Frolov},\ and\ \citenamefont
  {Kubiz\v{n}\'ak}}]{KrtousFrolovKubiznak:2008}%
  \BibitemOpen
  \bibfield  {author} {\bibinfo {author} {\bibfnamefont {Pavel}\ \bibnamefont
  {Krtou{\v s}}}, \bibinfo {author} {\bibfnamefont {Valeri~P.}\ \bibnamefont
  {Frolov}}, \ and\ \bibinfo {author} {\bibfnamefont {David}\ \bibnamefont
  {Kubiz\v{n}\'ak}},\ }\bibfield  {title} {\enquote {\bibinfo {title} {Hidden
  symmetries of higher dimensional black holes and uniqueness of the
  {Kerr-NUT-(A)dS} spacetime},}\ }\href@noop {} {\bibfield  {journal} {\bibinfo
   {journal} {Phys. Rev. D}\ }\textbf {\bibinfo {volume} {78}},\ \bibinfo
  {pages} {064022} (\bibinfo {year} {2008})},\ \Eprint
  {http://arxiv.org/abs/arXiv: 0804.4705 [hep-th]} {arXiv: 0804.4705 [hep-th]}
  \BibitemShut {NoStop}%
\bibitem [{\citenamefont {Houri}\ \emph {et~al.}(2008)\citenamefont {Houri},
  \citenamefont {Oota},\ and\ \citenamefont {Yasui}}]{HouriEtal:2008b}%
  \BibitemOpen
  \bibfield  {author} {\bibinfo {author} {\bibfnamefont {Tsuyoshi}\
  \bibnamefont {Houri}}, \bibinfo {author} {\bibfnamefont {Takeshi}\
  \bibnamefont {Oota}}, \ and\ \bibinfo {author} {\bibfnamefont {Yukinori}\
  \bibnamefont {Yasui}},\ }\bibfield  {title} {\enquote {\bibinfo {title}
  {Generalized {Kerr-NUT-de Sitter} metrics in all dimensions},}\ }\href
  {\doibase 10.1016/j.physletb.2008.07.075} {\bibfield  {journal} {\bibinfo
  {journal} {Phys. Lett.}\ }\textbf {\bibinfo {volume} {B666}},\ \bibinfo
  {pages} {391--394} (\bibinfo {year} {2008})},\ \Eprint
  {http://arxiv.org/abs/arXiv: 0805.0838 [hep-th]} {arXiv: 0805.0838 [hep-th]}
  \BibitemShut {NoStop}%
\bibitem [{\citenamefont {Houri}\ \emph {et~al.}(2009)\citenamefont {Houri},
  \citenamefont {Oota},\ and\ \citenamefont {Yasui}}]{HouriEtal:2009}%
  \BibitemOpen
  \bibfield  {author} {\bibinfo {author} {\bibfnamefont {Tsuyoshi}\
  \bibnamefont {Houri}}, \bibinfo {author} {\bibfnamefont {Takeshi}\
  \bibnamefont {Oota}}, \ and\ \bibinfo {author} {\bibfnamefont {Yukinori}\
  \bibnamefont {Yasui}},\ }\bibfield  {title} {\enquote {\bibinfo {title}
  {{Closed conformal Killing-Yano tensor and uniqueness of generalized
  Kerr-NUT-de Sitter spacetime}},}\ }\href {\doibase
  10.1088/0264-9381/26/4/045015} {\bibfield  {journal} {\bibinfo  {journal}
  {Class. Quant. Grav.}\ }\textbf {\bibinfo {volume} {26}},\ \bibinfo {pages}
  {045015} (\bibinfo {year} {2009})},\ \Eprint {http://arxiv.org/abs/arXiv:
  0805.3877 [hep-th]} {arXiv: 0805.3877 [hep-th]} \BibitemShut {NoStop}%
\bibitem [{\citenamefont {Chervonyi}\ and\ \citenamefont
  {Lunin}(2015)}]{ChervonyiLunin:2015}%
  \BibitemOpen
  \bibfield  {author} {\bibinfo {author} {\bibfnamefont {Yuri}\ \bibnamefont
  {Chervonyi}}\ and\ \bibinfo {author} {\bibfnamefont {Oleg}\ \bibnamefont
  {Lunin}},\ }\bibfield  {title} {\enquote {\bibinfo {title} {{Killing(-Yano)}
  tensors in string theory},}\ }\href@noop {} {\  (\bibinfo {year} {2015})},\
  \Eprint {http://arxiv.org/abs/arXiv: 1505.06154 [hep-th]} {arXiv: 1505.06154
  [hep-th]} \BibitemShut {NoStop}%
\bibitem [{\citenamefont {Houri}\ \emph {et~al.}(2012)\citenamefont {Houri},
  \citenamefont {Kubiz\v{n}\'{a}k}, \citenamefont {Warnick},\ and\
  \citenamefont {Yasui}}]{HouriEtal:2012}%
  \BibitemOpen
  \bibfield  {author} {\bibinfo {author} {\bibfnamefont {Tsuyoshi}\
  \bibnamefont {Houri}}, \bibinfo {author} {\bibfnamefont {David}\ \bibnamefont
  {Kubiz\v{n}\'{a}k}}, \bibinfo {author} {\bibfnamefont {Claude~M.}\
  \bibnamefont {Warnick}}, \ and\ \bibinfo {author} {\bibfnamefont {Yukinori}\
  \bibnamefont {Yasui}},\ }\bibfield  {title} {\enquote {\bibinfo {title}
  {Local metrics admitting a principal {Killing–Yano} tensor with torsion},}\
  }\href {http://stacks.iop.org/0264-9381/29/i=16/a=165001} {\bibfield
  {journal} {\bibinfo  {journal} {Class. Quantum Grav.}\ }\textbf {\bibinfo
  {volume} {29}},\ \bibinfo {pages} {165001} (\bibinfo {year} {2012})},\
  \Eprint {http://arxiv.org/abs/arXiv: 1203.0393 [hep-th]} {arXiv: 1203.0393
  [hep-th]} \BibitemShut {NoStop}%
\bibitem [{\citenamefont {Krtou{\v s}}\ \emph
  {et~al.}(2015{\natexlab{a}})\citenamefont {Krtou{\v s}}, \citenamefont
  {Kubiz\v{n}\'ak}, \citenamefont {Frolov},\ and\ \citenamefont
  {Kol\'{a}\v{r}}}]{KrtousEtal:2015}%
  \BibitemOpen
  \bibfield  {author} {\bibinfo {author} {\bibfnamefont {Pavel}\ \bibnamefont
  {Krtou{\v s}}}, \bibinfo {author} {\bibfnamefont {David}\ \bibnamefont
  {Kubiz\v{n}\'ak}}, \bibinfo {author} {\bibfnamefont {Valeri~P.}\ \bibnamefont
  {Frolov}}, \ and\ \bibinfo {author} {\bibfnamefont {Ivan}\ \bibnamefont
  {Kol\'{a}\v{r}}},\ }\bibfield  {title} {\enquote {\bibinfo {title} {Deformed
  and twisted black holes with {NUTs}},}\ }\href@noop {} {\  (\bibinfo {year}
  {2015}{\natexlab{a}})},\ \bibinfo {note} {to be published},\ \Eprint 
  {http://arxiv.org/abs/arXiv: 1511.02536 [hep-th]}
  {arXiv: 1511.02536 [hep-th]} \BibitemShut {NoStop}%
\bibitem [{\citenamefont {Krtou{\v s}}\ \emph
  {et~al.}(2015{\natexlab{b}})\citenamefont {Krtou{\v s}}, \citenamefont
  {Kubiz\v{n}\'ak},\ and\ \citenamefont
  {Kol\'{a}\v{r}}}]{KrtousKubiznakKolar:2015}%
  \BibitemOpen
  \bibfield  {author} {\bibinfo {author} {\bibfnamefont {Pavel}\ \bibnamefont
  {Krtou{\v s}}}, \bibinfo {author} {\bibfnamefont {David}\ \bibnamefont
  {Kubiz\v{n}\'ak}}, \ and\ \bibinfo {author} {\bibfnamefont {Ivan}\
  \bibnamefont {Kol\'{a}\v{r}}},\ }\bibfield  {title} {\enquote {\bibinfo
  {title} {{Killing--Yano} forms and {Killing} tensors on a warped space},}\
  }\href@noop {} {\  (\bibinfo {year} {2015}{\natexlab{b}})},\ \bibinfo {note}
  {to be published},\ \Eprint {http://arxiv.org/abs/arXiv: 1508.02642 [gr-qc]}
  {arXiv: 1508.02642 [gr-qc]} \BibitemShut {NoStop}%
\bibitem [{\citenamefont {Hinoue}\ \emph {et~al.}(2014)\citenamefont {Hinoue},
  \citenamefont {Houri}, \citenamefont {Rugina},\ and\ \citenamefont
  {Yasui}}]{HinoueEtal:2014}%
  \BibitemOpen
  \bibfield  {author} {\bibinfo {author} {\bibfnamefont {Kazuki}\ \bibnamefont
  {Hinoue}}, \bibinfo {author} {\bibfnamefont {Tsuyoshi}\ \bibnamefont
  {Houri}}, \bibinfo {author} {\bibfnamefont {Christina}\ \bibnamefont
  {Rugina}}, \ and\ \bibinfo {author} {\bibfnamefont {Yukinori}\ \bibnamefont
  {Yasui}},\ }\bibfield  {title} {\enquote {\bibinfo {title} {General wahlquist
  metrics in all dimensions},}\ }\href {\doibase 10.1103/PhysRevD.90.024037}
  {\bibfield  {journal} {\bibinfo  {journal} {Phys. Rev. D}\ }\textbf {\bibinfo
  {volume} {90}},\ \bibinfo {pages} {024037} (\bibinfo {year} {2014})},\
  \Eprint {http://arxiv.org/abs/arXiv: 1402.6904 [gr-qc]} {arXiv: 1402.6904
  [gr-qc]} \BibitemShut {NoStop}%
\bibitem [{\citenamefont {Kol\'a\v{r}}\ and\ \citenamefont {Krtou{\v
  s}}(2015)}]{KolarKrtous:2015}%
  \BibitemOpen
  \bibfield  {author} {\bibinfo {author} {\bibfnamefont {Ivan}\ \bibnamefont
  {Kol\'a\v{r}}}\ and\ \bibinfo {author} {\bibfnamefont {Pavel}\ \bibnamefont
  {Krtou{\v s}}},\ }\bibfield  {title} {\enquote {\bibinfo {title} {Weak
  electromagnetic field admitting integrability in {Kerr-NUT-(A)dS}
  spacetimes},}\ }\href {\doibase 10.1103/PhysRevD.91.124045} {\bibfield
  {journal} {\bibinfo  {journal} {Phys. Rev. D}\ }\textbf {\bibinfo {volume}
  {91}},\ \bibinfo {pages} {124045} (\bibinfo {year} {2015})},\ \Eprint
  {http://arxiv.org/abs/arXiv: 1504.00524 [gr-qc]} {arXiv: 1504.00524 [gr-qc]}
  \BibitemShut {NoStop}%
\bibitem [{\citenamefont {Kubiz\v{n}\'{a}k}(2009)}]{Kubiznak:2009}%
  \BibitemOpen
  \bibfield  {author} {\bibinfo {author} {\bibfnamefont {David}\ \bibnamefont
  {Kubiz\v{n}\'{a}k}},\ }\bibfield  {title} {\enquote {\bibinfo {title} {On the
  supersymmetric limit of {Kerr-NUT-AdS} metrics},}\ }\href@noop {} {\bibfield
  {journal} {\bibinfo  {journal} {Phys. Lett.}\ }\textbf {\bibinfo {volume}
  {B675}},\ \bibinfo {pages} {110--115} (\bibinfo {year} {2009})},\ \Eprint
  {http://arxiv.org/abs/arXiv: 0902.1999 [hep-th]} {arXiv: 0902.1999 [hep-th]}
  \BibitemShut {NoStop}%
\bibitem [{\citenamefont {Teitelboim}(1983)}]{Teitelboim:1983}%
  \BibitemOpen
  \bibfield  {author} {\bibinfo {author} {\bibfnamefont {Claudio}\ \bibnamefont
  {Teitelboim}},\ }\bibfield  {title} {\enquote {\bibinfo {title} {{Gravitation
  and hamiltonian structure in two spacetime dimensions}},}\ }\href {\doibase
  10.1016/0370-2693(83)90012-6} {\bibfield  {journal} {\bibinfo  {journal}
  {Phys. Lett.}\ }\textbf {\bibinfo {volume} {B126}},\ \bibinfo {pages}
  {41--45} (\bibinfo {year} {1983})}\BibitemShut {NoStop}%
\bibitem [{\citenamefont {Krtou{\v s}}(2007)}]{Krtous:2007}%
  \BibitemOpen
  \bibfield  {author} {\bibinfo {author} {\bibfnamefont {Pavel}\ \bibnamefont
  {Krtou{\v s}}},\ }\bibfield  {title} {\enquote {\bibinfo {title}
  {Electromagnetic field in higher-dimensional black-hole spacetimes},}\
  }\href@noop {} {\bibfield  {journal} {\bibinfo  {journal} {Phys. Rev. D}\
  }\textbf {\bibinfo {volume} {76}},\ \bibinfo {pages} {084035} (\bibinfo
  {year} {2007})},\ \Eprint {http://arxiv.org/abs/arXiv: 0707.0002 [hep-th]}
  {arXiv: 0707.0002 [hep-th]} \BibitemShut {NoStop}%
\end{thebibliography}


\end{document}